\documentclass[12pt]{article}

\usepackage{hyperref}       
\usepackage{url}            
\usepackage{booktabs}       
\usepackage{amsfonts}       
\usepackage{nicefrac}       
\usepackage{microtype}      
\usepackage{lipsum}
\usepackage{natbib}
\usepackage{amsmath}
\usepackage{graphicx,subfig}
\usepackage[margin=1.25in]{geometry}
\usepackage{mathtools}
\usepackage{authblk}
\usepackage{bm, bbm}
\usepackage{setspace}

\begin{document}

\onehalfspacing

\title{Bayesian non-parametric ordinal regression under a monotonicity constraint}

\author[1]{Olli~Saarela\thanks{Correspondence to: Olli Saarela, Dalla Lana School of Public Health, 155 College Street, Toronto, Ontario M5T 3M7, Canada. Email: \texttt{olli.saarela@utoronto.ca}}}
\author[2]{Christian~Rohrbeck}
\author[3,4]{Elja~Arjas}

\affil[1]{Dalla Lana School of Public Health, University of Toronto}
\affil[2]{Department of Mathematical Sciences, University of Bath}
\affil[3]{University of Helsinki}
\affil[4]{University of Oslo}
\date{\vspace{-2cm}}

\maketitle

\begin{abstract}
Compared to the nominal scale, the ordinal scale for a categorical outcome variable has the property of making a monotonicity assumption for the covariate effects meaningful. This assumption is encoded in the commonly used proportional odds model, but there it is combined with other parametric assumptions such as linearity and additivity. Herein, the considered models are non-parametric and the only condition imposed is that the effects of the covariates on the outcome categories are stochastically monotone according to the ordinal scale. We are not aware of the existence of other comparable multivariable models that would be suitable for inference purposes. We generalize our previously proposed Bayesian monotonic multivariable regression model to ordinal outcomes, and propose an estimation procedure based on reversible jump Markov chain Monte Carlo. The model is based on a marked point process construction, which allows it to approximate arbitrary monotonic regression function shapes, and has a built-in covariate selection property. We study the performance of the proposed approach through extensive simulation studies, and demonstrate its practical application in two real data examples.

\end{abstract}

\section{Introduction}\label{section:intro}

Ordinal data consist of categorical variables measured on a scale that has a natural ordering, but where there may not exist well-defined distances between the categories or where such distances have been left unspecified \citep{agresti2010analysis}. In regression models for ordinal data, the numbered categories of the response variable $Y \in \{1, \ldots, K\}$, $K > 1$ reflect a linear order of the possible response levels, while the predictors $\mathbf x = (x_1, \ldots, x_p)$ can be of any scale. In typical applications, such as when filling in questionnaires for political opinion polls, the categories could vary from $1$ signifying strong disagreement to $5$ representing strong agreement. Similarly, in customer rating of hotels or restaurants, the categories could reflect grades from a single star (*) to five stars (*****). Data of this type are commonly treated by applying the Likert scale with numerical values from 1 to 5, in spite of that it may be difficult to argue, for example, that the difference between grades 1 and 2 should be qualitatively similar to the difference between 4 and 5.

Compared to nominal scale, an important property of ordinal scale outcome variables is that for them the definition of monotonicity w.r.t.~the covariates is meaningful. This property was formulated by \citet{lehmann1966some} as positive regression dependency between variables $Y$ and $X$, meaning that the probabilities $P(Y \le y \mid X = x)$ are non-decreasing in $x$ \citep[][p. 43]{agresti2010analysis}. In the present multivariable ordinal regression context, the corresponding property can be defined as 
\begin{equation}\label{eq:mono}
\mathbf x_1 \le \mathbf x_2 \Rightarrow
S(k \mid \mathbf x_1) \le S(k \mid \mathbf x_2),
\end{equation}
where $S(k\mid \mathbf x)$ is the ``survival function'' $S(k\mid \mathbf x) = P(Y \ge k  \mid \mathbf x),~k=1, \ldots, K$, and $\mathbf x_1 \preceq \mathbf x_2$ is taken to mean that $x_{j1} \le x_{j2}$ for each covariate coordinate $j = 1, \ldots p$. We note that alternative definitions of multivariable monotonicity exist \citep[e.g.][]{fang2019multivariate}, but we will use \eqref{eq:mono} throughout.

Parametric forms of ordinal regression models have been considered widely in the statistical literature, with textbook-level treatments given, for example, by \citet{congdon2005bayesian}, \citet{johnson2006ordinal}, \citet{agresti2010analysis} and \citet{harrell2015regression}, and methodological reviews provided by \citet{ananth1997regression} and \citet{lall2002review}. By far the most popular model is the ordinal logistic, or proportional odds (PO) model, attributed to \citet{mccullagh1980regression}. The proportionality property is formulated by writing, for the response categories $k=2,\ldots,K$, the log-odds in the form of the linear expression
\begin{align}\label{eq:pom}
\log\left\{ \frac{S(k \mid \mathbf x)}
{1-S(k \mid \mathbf x)}\right\}
= \textrm{logit} \left\{S(k \mid \mathbf x)\right\} = \alpha_k + \mathbf \beta' \mathbf x
\end{align}
or, equivalently,
\begin{align*}
S(k \mid \mathbf x) = \textrm{expit}(\alpha_k + \beta' \mathbf x) = \frac{1}{1 + \exp\{-(\alpha_k + \beta' \mathbf x)\}}. 
\end{align*}
 
Since the values of the probabilities  $S(k\mid \mathbf x)$ decrease in $k$ for fixed $\mathbf x$, we have  $\alpha_2 > \alpha_3 > \ldots >\alpha_K$. The attribute ``proportional odds'' stems from that, in \eqref{eq:pom}, the same regression coefficients $\mathbf \beta$ are used for all $k = 2, \ldots, K$. Thus, in a comparison between  probabilities $ S(k\mid \mathbf x)$ and $ S(\ell\mid \mathbf x)$, when both are based on a common covariate value $\mathbf x$, the resulting odds ratio $\exp\{\alpha_k - \alpha_\ell\}$ is constant in $\mathbf x$. On the other hand, when fixing the category level $k$ but considering two covariate values $\mathbf x_1$ and $\mathbf x_2$, the odds ratio becomes $\exp \{ \mathbf \beta'(\mathbf x_1 - \mathbf x_2)\}$. Since this expression does not depend on $k$, it follows that, if the regression coefficients $\mathbf \beta_{j}$ are non-negative, then \eqref{eq:mono} holds for all $k=1,\ldots,K$. Expressed in words, increasing the values of the covariates $\mathbf x$ has the effect of stochastically increasing the outcome variable $Y$ with respect to the ordering $\prec$. The conclusion on the right of \eqref{eq:mono} can be denoted compactly by $Y \mid \mathbf x_1 \prec_{\textrm{st}} Y \mid \mathbf x_2$. It is obvious that if $\beta_j<0$ for some $j=1,\ldots,p$, the conclusion still holds if $\mathbf \beta'(\mathbf x_1 - \mathbf x_2) > 0$.  \citet{espinosa2019constrained} recently extended the PO formulation to include ordinal predictors, entered into the linear predictor through dummy variables, with monotonicity constraints for the respective regression coefficients.

Several authors have noted that the constant proportionality assumption of model~\eqref{eq:pom} can in practice be unduly restrictive in order to provide an adequate description for empirical data. Statistical tests for assessing the validity of the proportionality  property have  been derived, either by assessing the general goodness-of-fit of the model \citep{ashby1986ordered} or by comparing it to models in which  this property has been relaxed \citep{brant1990assessing}. In the non-proportional odds (NPO) model, the regression coefficients $\beta_{j}$ are allowed to vary with category level $k$, while in the partial proportional odds (PPO) model, such variation is possible in a subset of all levels \citep[e.g.][]{peterson1990partial,bender1998using,tutz2003ordinal}. The key challenge in fitting such models is to ensure that the stochastic ordering conditions hold for any chosen explanatory variables  \citep{mckinley2015bayesian}. Additional variants consist of cumulative link models \citep{agresti2003categorical}, where the link function differs from the logit, common choices being probit, log-log and complementary log-log.

Bayesian formulations of the PO model are commonly based on introducing a model for a continuous-valued latent variable $y^* = \beta'\mathbf x + \varepsilon$, where $y = k$ if $\alpha_k < y^* \le \alpha_{k+1}$, or equivalently $P(Y=k \mid \mathbf X)
= P(\alpha_k - \beta' \mathbf x < \varepsilon < \alpha_{k+1} - \beta' \mathbf x)$. Assuming $\varepsilon$ to be standard logistic or standard normally distributed would give the logit and probit models, respectively; in the latter, computational efficiency has been enhanced by applying data augmentation for the latent variable  \citep[e.g.][]{albert1993bayesian,johnson2006ordinal}. The latent variable formulation has also enabled further semi-parametric and non-parametric extensions of the original model, building on \citet{kottas2005nonparametric}. Rather than relying on a distributional assumption such as $\varepsilon \mid \phi \sim \textrm{Logistic}(0,\phi)$ or $\varepsilon \mid \phi \sim N(0,\phi)$, \citet{chib2010additive} generalized the model to allow for Dirichlet process normal mixture errors, taking $\phi^{-1} \mid G \sim G$ and $G \sim \textrm{DP}(a, G_0)$, where $G_0 = \textrm{Gamma}(\phi^{-1} \mid .)$. This construction can be understood as allowing for a flexible link function instead of the logit or probit links implied by the common distributional assumptions. In addition to the linear covariate effects (or effects of binary covariates), \citet{chib2010additive} extended the model to a GAM-type formulation $y^* = \beta'\mathbf x + \sum_{j=1}^q g_j(z_j) + \varepsilon$, where they modeled the functions flexibly through a cubic spline construction.

Latent variable formulations are relatively straightforward to extend to multivariate ordinal outcomes as done in \citet{bao2015bayesian} and \citet{deyoreo2018bayesian}. \citet{deyoreo2018bayesian} proposed a non-monotonic density regression based approach for modeling the joint effects of multiple covariates. Here the Dirichlet process normal mixture model is assumed jointly for the potentially multivariate latent outcome $\mathbf y^*$ and covariate vector $\mathbf x$ so that $(\mathbf y^*, \mathbf x) \mid (\boldsymbol \mu, \boldsymbol \Psi) \sim N(\boldsymbol \mu, \boldsymbol \Psi)$, $(\boldsymbol \mu, \boldsymbol \Psi) \mid G \sim G$, and $G \sim \textrm{DP}(a, G_0)$, with the base distribution $G_0 = N(\boldsymbol \mu \mid .)IW(\boldsymbol \Psi \mid .)$. Marginal regression curves or surfaces may then be calculated from posterior samples from the joint distribution. We discuss possible extension of our proposal to multivariate responses briefly in Section \ref{sec:discussion}, but our main focus herein is in multivariable regression surfaces.

In a slightly different context, latent variable formulations have also been considered in item response theory, combined with a monotonicity assumption. Although the models therein are usually for multiple response items (assumed independent conditionally on the latent trait), the monotonicity is usually formulated with respect to a unidimensional latent trait being measured \citep{van2001relationships,van2007mokken}, whereas herein we consider multivariable monotonicity with respect to multiple observed covariates. Instead of assuming monotonicity, in item response models it is sometimes of interest to test the monotonicity, as demonstrating the ordering of the response categories is important for the item to be used as part of a measurement instrument. \citet{luzardo2015nonparametric} presented a non-parametric estimator for a monotonic item response model for a multidimensional latent trait based on kernel density regression, but only considered dichotomous item responses.

The practical goal of the statistical literature briefly reviewed above has been in attempts to explain and to quantify the effects which variations in the covariate values have on the outcome of interest.  Ordinal regression has been studied extensively also in the machine learning literature, albeit with a somewhat different focus, from the perspective of successful automated classification. Inference has then had a smaller or no role. A good survey is provided by \citet{gutierrez2015ordinal}. 

In the present paper, our interest is in non-parametric estimation under the monotonicity constraint \eqref{eq:mono}, in a form that would not impose additional parametric or distributional assumptions or make use of continuous latent variable formulations. Our goal is then to demonstrate that, by Bayesian `borrowing of strength' from neighbouring model structures, the constraint would in itself impose a sufficient degree of regularity for successful estimation of the regression functions. We note that in the absence of the PO assumption, relatively large sample sizes are required for non-parametric multivariable regression.

The corresponding non-parametric constrained optimization problem of minimizing  the sum of squared prediction errors, under monotonicity constraints, is known as isotonic regression \citep[see][and the references therein]{barlow1972isotonic}. \citet{kotlowski2012nonparametric} considered an extension of this for the classification of an ordinal response variable under \eqref{eq:mono}. Their approach directly minimized a linear loss function of the difference between the actual and predicted categories. \citet{stout2015isotonic} proposed an algorithm for multivariable isotonic regression. However, we are not aware of the existence of non-parametric and monotonic multivariable ordinal regression approaches that could be used for inferential purposes. To achieve this, we extend in this work the Bayesian monotonic regression approach of \citet{saarela2011method} for ordered multi-category responses. 

We note that there are other Bayesian monotonic regression formulations that could potentially be extended to ordinal responses, such as the projection-based approach of \citet{lin2014bayesian} or the tree-based one of \citet{chipman2016high}. We chose the approach of \citet{saarela2011method} for the extension since it can approximate arbitrary monotonic regression functions, it incorporates a useful covariate selection feature, handles both continuous and ordinal predictors, and can be naturally adapted to multiple ordered regression surfaces without additional structural assumptions. The formulation is based on a marked point process prior, and the estimation procedure is based on reversible jump Markov chain Monte Carlo  \citep{green1995reversible}. 

The paper is structured as follows: in Section \ref{sec:model} we formulate the proposed model and MCMC algorithm for estimation. In Section \ref{sec:sim} we present results of simulation studies for the properties of the method, and in Section \ref{sec:data} we demonstrate its use for real data. We conclude with a discussion in Section \ref{sec:discussion}.

\section{Model and estimation method}\label{sec:model}

\subsection{Model construction}\label{sec:constr}

Our notation for marked point processes broadly follows that of \citet{moller2003statistical}. We introduce spatial point processes $\Xi_i$, $i = 1, \ldots, s,$ as random countable subsets of bounded spaces $\mathcal S_i \subseteq \mathbb R^p$, with finite realizations of $n(\Xi_i)$ points. The numbered point coordinates of a realization are denoted by $\boldsymbol \xi_{ij} \in \mathcal S_i$, $j = 1, \ldots, n(\Xi_i)$. To each point we can attach a random variable, mark $\delta_{ij} \in \mathcal A \subseteq \mathbb R$, to specify a marked point process $\Delta_i = \{(\boldsymbol \xi_{ij}, \delta_{ij})\} \subset \mathcal S_i \times \mathcal A$, with $n(\Delta_i) = n(\Xi_i)$. 

In the multivariable regression context we could choose $s=1$ and then let the space to be $\mathcal S_1 = [0,1]^p$, the $p$-cube spanned by the $p$ covariate axes, each scaled to the interval $[0,1]$. This is the domain of the regression function being modeled. The marks in turn represent the levels of the regression function at the points, and in the absence of a link function, we can take $\mathcal A = [0,1]$ for modeling a dichotomous response variable. However, to enable covariate selection functionality, \citet{saarela2011method} proposed also specifying the lower dimensional point processes to allow dropping some of the covariate axes from the model. Thus, the model can make use of all $s=2^p-1$ point processes corresponding to the non-empty subsets of $\{1, \ldots, p\}$. Due to the need to place the points in the common covariate space $[0,1]^p$, to construct the regression surfaces, we therefore introduce the notation for completed point coordinates $\boldsymbol{\tilde \xi}_{ij} \in [0,1]^p$, where the `missing' coordinates are set to zero. For example, with $p=2$, we can take $s=3$ with $\mathcal S_1 = \mathcal S_2 = [0,1]$ and $\mathcal S_3 = [0,1]^2$. Then $\boldsymbol{\tilde \xi}_{1j} = (\xi_{1j1}, 0)$, $\boldsymbol{\tilde \xi}_{2j} = (0, \xi_{2j2})$ and $\boldsymbol{\tilde \xi}_{3j} = (\xi_{3j1}, \xi_{3j2}) \in [0,1]^2$. If, for example, $n(\Delta_2) = n(\Delta_3) = 0$ and $n(\Delta_1) > 0$, covariate $X_2$ is not selected in the resulting regression function realization.

\citet{saarela2011method} proposed monotonic regression modeling based on such a marked point process construction for conditional expectations $E[Y \mid \mathbf x; \lambda]$, where $\lambda(\mathbf x)$ is a realization of a random monotonic function of the covariates $\mathbf x$. This is applicable to modeling continuous outcomes with an additional distributional assumption or to dichotomous outcomes without further assumptions. To extend such a model to ordinal responses, we introduce functions $S(k \mid \mathbf x; \lambda_k) = P(Y \ge k \mid \mathbf x; \lambda_k)$, $k = 2, \ldots, K$, where $\lambda_k : [0,1]^p \rightarrow \mathcal A$ is a realization of a random regression function mapping the covariate space to probabilities of the outcome variable categories, possibly on a link function scale. We formulate the model first without a link function, in which case we have simply that $\mathcal A = [0,1]$ and $S(k \mid \mathbf x; \lambda_k) = \lambda_k(\mathbf x)$. The likelihood for the $\lambda_k$s, given the observed data $\mathcal D_N=\{(\mathbf x_n,Y_n); 1 \le n \le N\},$ is then
\begin{align*}
\mathcal L(\lambda_1, \ldots, \lambda_K; \mathcal D_N) = 
\prod_{n=1}^N \prod_{k=1}^K
[S(k \mid \mathbf x_n; \lambda_k) - S(k+1 \mid \mathbf x_n; \lambda_{k+1})]^{\mathbf 1_{\{Y_n = k\}}},
\end{align*}
where $S(1 \mid \mathbf x) = 1$ and $S(K+1 \mid \mathbf x) = 0$ by definition. Our interest is in the posterior distribution of the $\lambda_k$s under a suitably flexible prior specification. The key monotonicity property postulated is that, for $1 \le k \le K$, the realizations $S(k \mid \mathbf x; \lambda_k)$  are non-decreasing in $\mathbf x$, i.e., $S(k \mid \mathbf x_1; \lambda_k) \le S(k \mid \mathbf x_2; \lambda_k)$ whenever $\mathbf x_1 \preceq \mathbf x_2$. This is without loss of generality, because if the direction of monotonicity along some covariate axis goes to the opposite direction, the original covariate $x_j$ scaled to the interval $[0,1]$ can be transformed simply to $x_j' = 1 - x_j$ when entered into the model. In fact, in Section \ref{sec:direction} we propose an extension where the direction of monotonicity does not need to be fixed a priori, but rather is specified as a random variable and selected data-adaptively as part of the estimation procedure. In addition to the monotonicity property, the `survival' probabilities of ordered outcome categories are restricted by $S(k \mid \mathbf x; \lambda_k) \ge S(\ell \mid \mathbf x; \lambda_k)$, where $1 \le k < \ell \le K$, for all $\mathbf x$. The same ordering constraints apply obviously to the $\lambda_k$s also if specified on the scale of a monotonic link function. 

When extending the previously outlined marked point process construction to ordinal regression, we take $\Delta_i = \{(\boldsymbol \xi_{ij}, \boldsymbol \delta_{ij})\} \subset \mathcal S_i \times [0,1]^{K-1}$, with each mark now a random vector $\boldsymbol \delta_{ij} = (\delta_{ij1}, \ldots, \delta_{ijK})$, where $\delta_{ij1} \ge \delta_{ij2} \ge \ldots \ge \delta_{ijK}$, with $\delta_{ij1} = 1$ by definition, and $\delta_{ijk}$ reflecting the value of the regression function $\lambda_k(\mathbf x)$ at point $\mathbf x = \boldsymbol{\tilde \xi}_{ij}$, i.e., $\lambda_k(\boldsymbol{\tilde \xi}_{ij}) = \delta_{ijk}$. In addition, a fixed point $(\boldsymbol{\tilde \xi}_{01}, \boldsymbol{\delta}_{01})$ with mark $\boldsymbol \delta_{01} = (\delta_{011}=1, \delta_{012}, \ldots, \delta_{01K})$ is placed at the origin. It is obvious that the value of the regression function $\lambda_k(\mathbf x)$ at point $\mathbf x$ is constrained to lie in the interval $[\max\{\delta_{ijk} : \boldsymbol{\tilde \xi}_{ij} \preceq \mathbf x\},\min\{\delta_{ijk} : \boldsymbol{\tilde \xi}_{ij} \succeq \mathbf x\}$]. In principle there would be many alternative ways to define the regression function realizations with these properties, but for computational simplicity we choose $\lambda_k(\mathbf x) = \max\{\delta_{ijk} : \boldsymbol{\tilde \xi}_{ij} \preceq \mathbf x\}$, $1 \le k \le K$. The resulting piecewise constant regression function realizations enable efficient local updating moves in the estimation algorithm (Section \ref{section:mcmc}), while still allowing the construction to approximate general monotonic functions with arbitrary precision by increasing the number of support points (Section~S1 of the online Supplementary Material). The realizations are monotonic for each $k = 1, \ldots, K$ because $\mathbf x_1 \preceq \mathbf x_2$ $\Rightarrow$ $\{\delta_{ijk} : \boldsymbol{\tilde \xi}_{ij} \preceq \mathbf x_1\} \subseteq \{\delta_{ijk} : \boldsymbol{\tilde \xi}_{ij} \preceq \mathbf x_2\}$ $\Rightarrow$ $\max\{\delta_{ijk} : \boldsymbol{\tilde \xi}_{ij} \preceq \mathbf x_1\} \le \max\{\delta_{ijk} : \boldsymbol{\tilde \xi}_{ij} \preceq \mathbf x_2\}$ $\Rightarrow$ $\lambda_k(\mathbf x_1) \le \lambda_k(\mathbf x_2)$.

\begin{figure}[!h]
        \centering
        \includegraphics[width=\textwidth]{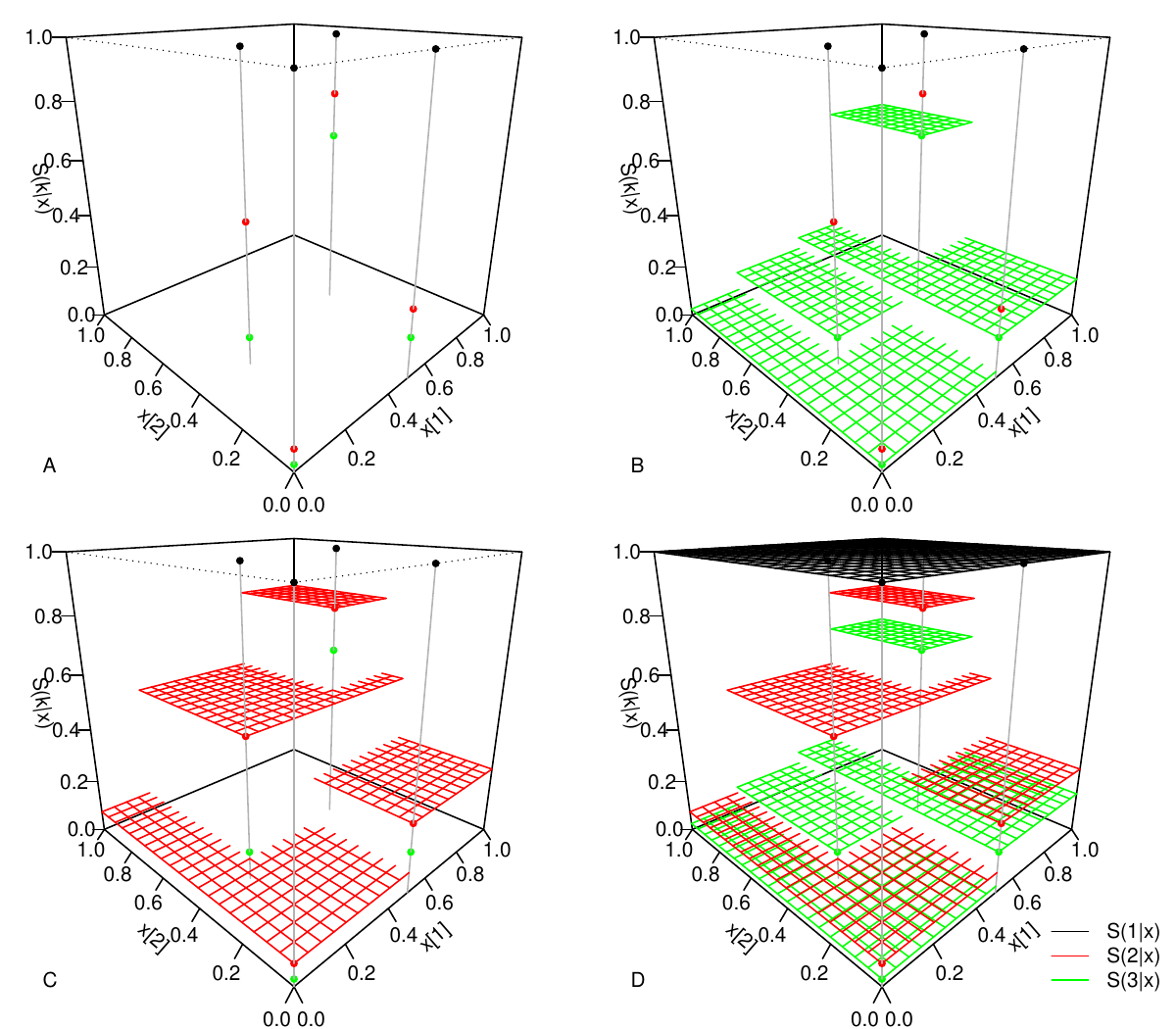}
        \caption{An example of a marked point process realization and the resulting regression function realizations. Panel A shows the marked point process realizations $\Delta_1 = \{(\boldsymbol \xi_{11}, \boldsymbol \delta_{11})\}$ and $\Delta_3 = \{(\boldsymbol \xi_{31}, \boldsymbol \delta_{31}), (\boldsymbol \xi_{32}, \boldsymbol \delta_{32})\}$. Panel B additionally shows the implied regression function realization $\lambda_3(\mathbf x)$, panel C shows $\lambda_2(\mathbf x)$ and panel D shows all $\lambda_i(\mathbf x)$, $i = 1, 2, 3$.} \label{schematic}
\end{figure}

For illustration, an example point and mark configuration with $p=2$, $s=3$ and $K=3$ is presented in Figure \ref{schematic}. Here $S_1 = S_2 = [0,1]$ and $S_3 = [0,1]^2$. The realization depicted in panel A involves the fixed mark at origin, with $\boldsymbol \delta_{01} = (\delta_{011}, \delta_{012}, \delta_{013}) = (1, 0.075, 0.025)$. In addition $n(\Delta_1) = 1$ with $(\boldsymbol {\tilde \xi}_{11}, \boldsymbol \delta_{11}) = ((0.5, 0.0), (1, 0.25, 0.15))$, $n(\Delta_2) = 0$ and $n(\Delta_3) = 2$ with
$(\boldsymbol {\tilde \xi}_{31}, \boldsymbol \delta_{31}) = ((0.2, 0.4), (1, 0.5, 0.1))$ and $(\boldsymbol {\tilde \xi}_{32}, \boldsymbol \delta_{32}) = ((0.7, 0.5), (1, 0.8, 0.65))$. Panels B-D in Figure \ref{schematic} show how the regression function realizations $\lambda_2(\mathbf x)$ and $\lambda_3(\mathbf x)$ are determined by the points and marks. The example shows how these simultaneously satisfy the monotonicity property along the covariate axes and the ordering of the survival functions of the outcome categories.

We note that in the proposed construction, the locations of the support points $\boldsymbol{\tilde \xi}_{ij}$ are shared between the $K-1$ regression functions, with separate ordered levels. Alternatively it would have been possible to define separate spatial point processes for the different functions, but the proposed structure is more parsimonious, while still allowing the levels $\delta_{ijk}$ to move freely within the ordering constraints.

While other prior specifications would be possible, herein each point process is a priori taken to be a homogeneous Poisson point process on $\mathcal S_i$ with rate $\rho_i \sim \textrm{Gamma}(a,b)$. Here we suggest choosing the shape and rate parameters $a=b$ to take a small value such as 0.1 to be relatively agnostic on how many points to use if the point process is used in the regression surface construction, but placing enough probability mass close to zero to allow the point process to be effectively removed from the construction to reduce dimension. The latter can occur when the rate parameter estimates are to close to zero, and there are no points in the realization for the particular process. Conditionally on the current point configuration, the marks are assumed a priori jointly uniformly distributed in the space restricted by the ordering constraints. Thus, the only specific prior information is expressed in the monotonicity assumption.

\subsection{Markov chain Monte Carlo algorithm}\label{section:mcmc}

We use MCMC to propose local modifications to the current regression function realizations, thereby exploring the space of possible versions of such functions, while simultaneously enabling fully probabilistic inferences. Here the postulated ordering constraints provide information for the construction of reasonable proposals. While prior proposals may not generally provide sufficient information for constructing a well mixing chain, in the present setting they will in fact be very useful as the ordering constraints for the fully conditional priors are highly informative for making small local updates to the regression surfaces. The prior proposals also lead to simple forms of the  corresponding Metropolis-Hastings (M-H) ratios. To demonstrate this, consider a current point configuration with the total of $M = (K-1)(1 + \sum_{i=1}^s n(\Delta_i))$ marks. Denoting $\boldsymbol \delta_i = (\boldsymbol \delta_{i1}, \boldsymbol \delta_{i2}, \ldots, \boldsymbol \delta_{in(\Delta_i)})$, $i \in \{1, \ldots , s\}$, and $\boldsymbol \delta_0 = (\boldsymbol \delta_{01})$ for the fixed point at the origin, the joint prior density of all the marks $\boldsymbol \delta = (\boldsymbol \delta_0, \boldsymbol \delta_1, \ldots, \boldsymbol \delta_s)$
has the expression  
\begin{align*}
f(\boldsymbol \delta) = \frac{M!}{M^*} \left(\frac{1}{|\mathcal A|}\right)^M. \end{align*}
Here $M!$ is the total number of permutations, and $M^*$ is the number of these permutations that satisfy the ordering constraints depending, in part, on the locations of the points. All  resulting fully conditional distributions are also uniform; to see this, we can write, for example, the conditional density for the mark vector of a single point, given all others, as
\begin{align*}
f(\boldsymbol \delta_{ij} \mid
\boldsymbol \delta_{-ij}) = \frac{f(\boldsymbol \delta)}{f(\boldsymbol \delta_{-ij})}
= \frac{M!}{M^*} \left(\frac{1}{|\mathcal A|}\right)^M \frac{1}{f(\boldsymbol \delta_{-ij})}
\end{align*}
for values $\boldsymbol \delta_{ij}$ satisfying the ordering constraints, and zero otherwise. Here the marginal density in the denominator simply normalizes the uniform density in the numerator. The marginal densities are generally non-uniform but need not be known as the conditional priors can be simulated, and cancel out from the M-H ratios without having to evaluate the normalizing constant. In practice this can proceed by sampling uniformly distributed mark vectors until one satisfying the ordering constraints is drawn.

Some of the proposals change the dimension of the parameter space, requiring reversible jump type updating moves for dimension matching \citep{green1995reversible}. The proposed moves in the algorithm and the corresponding M-H ratios, as well as other details of the implementation, are listed in Section~S2 of the online Supplementary Material. The proposals always satisfy the partial ordering constraints and thus produce monotonic realizations of the regression surfaces. We verified that the algorithm produces samples from the homogeneous Poisson point process priors when run without data.

\subsection{Extensions}

\subsubsection{Semi-parametric models}\label{sec:SemiparametricModel}

While the assumption of monotonicity provides structure to the problem, estimation of non-parametric models involving a large number of covariates eventually becomes difficult due to the curse of dimensionality. To be able to incorporate more covariates, and use the non-parametric monotonic structure where it is most needed, we introduce also a semi-parametric version of the model. We allow the functions $S(k \mid \mathbf x, \mathbf z; \lambda_k, \theta)$ to depend on additional covariates $\mathbf z = (z_1, \ldots, z_q)$ and parameters $\theta$, with the split of the covariates into $\mathbf x$ and $\mathbf z$ determined a priori. Using the logit link as in \eqref{eq:pom}, the regression functions $\lambda_k$ now take the role of the level-specific intercept terms in \eqref{eq:pom}:
\begin{equation}
\log\left( \frac{S(k \mid \mathbf x, \mathbf z; \lambda_k, \theta)}
{1-S(k \mid \mathbf x, \mathbf z; \lambda_k, \theta)}\right)
= \lambda_k(\mathbf x) + \sum_{j=1}^q \beta_j z_j,
\label{eq:ModelSemiparametric}
\end{equation}
where $\theta=(\beta_1, \ldots, \beta_q)$. The prior specification for $\lambda_k$s is otherwise the same as in Section \ref{sec:constr}, but instead of the interval $[0,1]$, these functions map to some a priori chosen interval $A \subset \mathbb R$.

The log-linear effects in \eqref{eq:ModelSemiparametric} can be further replaced by a GAM-type specification
\begin{equation}
\log\left( \frac{S(k \mid \mathbf x, \mathbf z; \lambda_k, \theta)}
{1-S(k \mid \mathbf x, \mathbf z; \lambda_k, \theta)}\right)
= \lambda_k(\mathbf x) + \sum_{j=1}^q g(z_j; \boldsymbol \beta_j),
\label{eq:ModelSemiparametric2}
\end{equation}
where $g(z_j; \boldsymbol \beta_j)$ is some flexible function of $z_j$ parametrized with $\boldsymbol \beta_j = (\beta_{j1}, \ldots, \beta_{jM})$. There would be many ways to specify such a parametrization, either monotonic or unrestricted, for example, by employing splines. We chose to use non-monotonic piecewise constant functions with sufficiently many equal length non-overlapping intervals $[c_1, c_2), \ldots, [c_M, c_{M+1})$, so that $g(z_j; \beta_j) = \sum_{m=1}^M \beta_{jm} \mathbf 1_{\{c_m \le z_j < c_{m+1}\}}$. If the number of intervals $M$ is large, in the absence of a monotonicity restriction some smoothing is required to estimate the coefficients; here we consider normal random walk priors, which in the first order case would be given by $\beta_{jm} \mid (\beta_{j(m-1)}, \ldots, \beta_{j1}) \sim N(\beta_{j(m-1)}, \phi_j)$, $m \ge 2$. An alternative would be the second order prior considered e.g., by \citet{berzuini1994bayesian}, which implies more smoothness. For identifiability, we could set $\beta_{j1} = 0$, or use an improper flat prior for the $\beta_{j1}$s together with a sum to zero constraint for each vector $\boldsymbol \beta_j$ over the observations. We chose the latter option as it does better in terms of the autocorrelation of the updates. In Section \ref{sec:data} we use versions of models \eqref{eq:ModelSemiparametric} and \eqref{eq:ModelSemiparametric2} where $\lambda_k(\mathbf x) = \lambda_k$ are just the ordered intercepts and $z_j = x_j$, $j = 1, \ldots, p=q$, as comparators to our model.

With a view to the data analysis example of Section \ref{sec:data}, note that we can deal with clustered data by introducing random effects into the linear predictor, such as in  
\begin{align}\label{eq:ModelSemiparametric3}
\log\left( \frac{S(k \mid \mathbf x, \mathbf z, c; \lambda_k, \theta)}
{1-S(k \mid \mathbf x, \mathbf z, c; \lambda_k, \theta)}\right)
= \lambda_k(\mathbf x) + \sum_{j=1}^q \beta_j z_j + \gamma_{c}.
\end{align}
Here $\theta=(\beta_1, \ldots, \beta_q, \gamma_1, \ldots, \gamma_C)$, $c \in \{1, \ldots, C\}$ is an observed variable indicating cluster membership, and the cluster effects (`random intercepts') are a priori independent and identically distributed draws from a distribution such as $\gamma_{c} \sim N(0, \tau^2)$. Updating the parameters in the models \eqref{eq:ModelSemiparametric}-\eqref{eq:ModelSemiparametric3} is detailed in Section~S2 of the online Supplementary Material.

\subsubsection{Models with unknown direction of monotonicity}\label{sec:direction}

In some situations, such as the example in Section \ref{sec:credit}, we may not want to assume a priori the direction of monotonicity, but still want to assume the regression function realizations to be multivariable monotonic to relax the PO
and linearity assumptions. In the proposed framework this can be incorporated into the model by choosing the direction of the monotonicity for covariate $j$ by entering it on the original scale (scaled to $[0,1]$) or inverting it as $x_j' = x_j^{\psi_j}(1-x_j)^{1-\psi_j}$, where the prior $\psi_j \sim \textrm{Bernoulli}(0.5)$ would reflect being agnostic about the direction of the covariate effect. In the MCMC algorithm, changes to $\psi_j$ can then be proposed jointly with a birth or combined death-birth step when in an empty point configuration, that is, when $n(\Delta_i) = 0$ for all the point processes $i$ specified in a space $\mathcal S_i$ involving the covariate axis $j$. When proposing the new direction with equal probabilities, the priors and proposal distributions cancel out from the M-H ratio, leaving the proposals to be accepted as specified for the birth and combined death-birth steps in Section~S2 of the online Supplementary Material. The likelihood is evaluated by building the regression surfaces $\lambda_k(\mathbf x')$, $k = 1, \ldots, K$, with the potentially inverted covariates $\mathbf x' = (x_1', \ldots, x_p')$.

\subsubsection{Conditional prior to counter ``spiking'' behaviour}\label{section:spiking}

The classical isotonic regression is reported to produce inconsistent estimates at the boundaries of the support of the data; this phenomenon was called the ``spiking'' problem by \citet{wu2015penalized}, who proposed penalized estimation as a solution. The unstable behaviour at the boundaries can be an issue also for the proposed Bayesian implementation as it does not impose any smoothness on the regression function realizations. To counter such behaviour close to the origin, where the function levels are not constrained from below by other support points, we also experimented with a modified prior, where the function levels $\delta_{01k}$, $k \in \{2, \ldots, K\}$, at the origin have the conditional prior specification $\delta_{01k} \mid \boldsymbol\delta_{-01k} \sim \textrm{Beta}(1+\min(\sum_{i=1}^s n(\Delta_i), d),1) \times (\delta_{\max} - \delta_{\min}) + \delta_{\min}$, that is, a left-skewed Beta distribution scaled to the interval $[\delta_{\min}, \delta_{\max}]$ determined by the partial ordering constraints imposed by the current point configuration. This prior allows borrowing information from the regression surface levels near the origin; the updating moves are as in Step 6 in Section~S2 of the online Supplementary Material, using prior proposals accepted with the likelihood ratio. Here the choice of the constant $d$ determines the level of penalization, with $d=0$ returning the previous non-informative uniform prior. 

Some results from applying this alternative prior, in the context of the discontinuous survival functions of Section 3.1, are described in Section~S3.4 of the online Supplementary Material.

\section{Numerical Examples}
\label{sec:sim}

We apply our approach to the two ordinal regression models described in Section~\ref{sec:model}. In Section~\ref{subsec:SimNonparametric}, the survival functions are defined directly in terms of non-parametrically specified monotonic functions, that is, $S(k\mid\mathbf{x};\lambda_k)=\lambda_k(\mathbf{x}), ~k=1,\ldots,K$. Section~\ref{subsec:SimSemiparametric} considers the semi-parametric model structure defined in (3). Across all studies, the number of categories was set to $K=5$, corresponding to the Likert scale with values from 1 to 5. The functions $\lambda_1,\ldots,\lambda_K$ were defined on the unit square, i.e., $\lambda_k:[0,1]^2\to\mathbb{R},~k=1,\ldots,5$.

In the experiments, data sets of size $N=1000$ and $N=5000$ were generated, with values $\mathbf{x}$ sampled independently from the uniform distribution on the unit square. The samples of size $N=1000$ were subsets of those of size $N=5000$, and 20 independent repetitions of this procedure were performed. Independent gamma priors were specified for the point process intensities, $\rho_i\sim\mbox{Gamma}(0.1,0.1), ~i=1,\ldots,s$. Approximate samples from the posterior were obtained by running the Markov chain Monte Carlo sampler, described in Section~\ref{sec:model}, for 500,000 iterations after a burn-in period of 100,000, and then saving every 50th state of the chain. Trace plots illustrating convergence and mixing of the functional levels $\lambda_k(\mathbf{x})$ at some of the simulated values of $\mathbf{x}$ are provided in Section~S3.2 of the Supplementary Material. 

To measure performance, we consider the probabilities $P(k\mid\mathbf{x}_n)$ instead of the monotonic functions $\lambda_k(\mathbf{x}_n), ~n=1,\ldots,N$. Let $\hat{p}(k\mid\mathbf{x}_n)$ denote the posterior mean probability of category $k$ at the covariate value $\mathbf{x}_n$, computed as the Monte Carlo average of the corresponding sampled values. Then, for each category, we calculate the mean absolute difference between the true probability $P(k\mid\mathbf{x}_n)$ and $\hat{p}(k\mid\mathbf{x}_n)$ across the data points with $y_n=k$. Formally, for $\mathcal{D}_k = \left\{n\in\left\{1,\ldots,N\right\}\,:\,y_n=k\right\}$ and $L$ posterior samples, the performance measure for category $k$ is defined as the mean absolute error
\begin{equation}
\mathrm{MAE}(k)=
\frac{1}{|\mathcal{D}_k|L}\sum_{n\in\mathcal{D}_k}\sum_{\ell=1}^L\left|\hat{p}^{(\ell)}\left(y_n\mid\mathbf{x}_n\right) - P\left(y_n\mid\mathbf{x}_n\right)\right|\label{eq:MAEk},
\end{equation}
where $\hat{p}^{(\ell)}$ denotes the estimated probability based on the $\ell$-th sample.  

Another aspect we wish to investigate is the ability of our approach to estimate the survival function $S(k\mid\cdot), ~k=1,\ldots,K$ across the covariate space, rather than just at the points $\mathbf{x}_n$ with $y_n=k$. Again, we focus on the observed covariate values, $\mathbf{x}_1,\ldots,\mathbf{x}_N$, but now consider the vector $[P(1\mid\mathbf{x}_n), \ldots,P(K\mid\mathbf{x}_n)]$ instead of $P(y_n\mid\mathbf{x}_n)$.  The overall model fit is then measured by
\[
\mathrm{MAE} = \frac{1}{NKL}\sum_{n=1}^N \sum_{k=1}^K \sum_{\ell=1}^L \left|\hat{p}^{(\ell)}(k\mid\mathbf{x}_n) - P(k\mid\mathbf{x}_n)\right|.
\]

\subsection{Non-parametric model structures}
\label{subsec:SimNonparametric}

\begin{figure}
\centering
\includegraphics[width=0.24\textwidth, trim={2cm 2cm 2cm 2cm}]{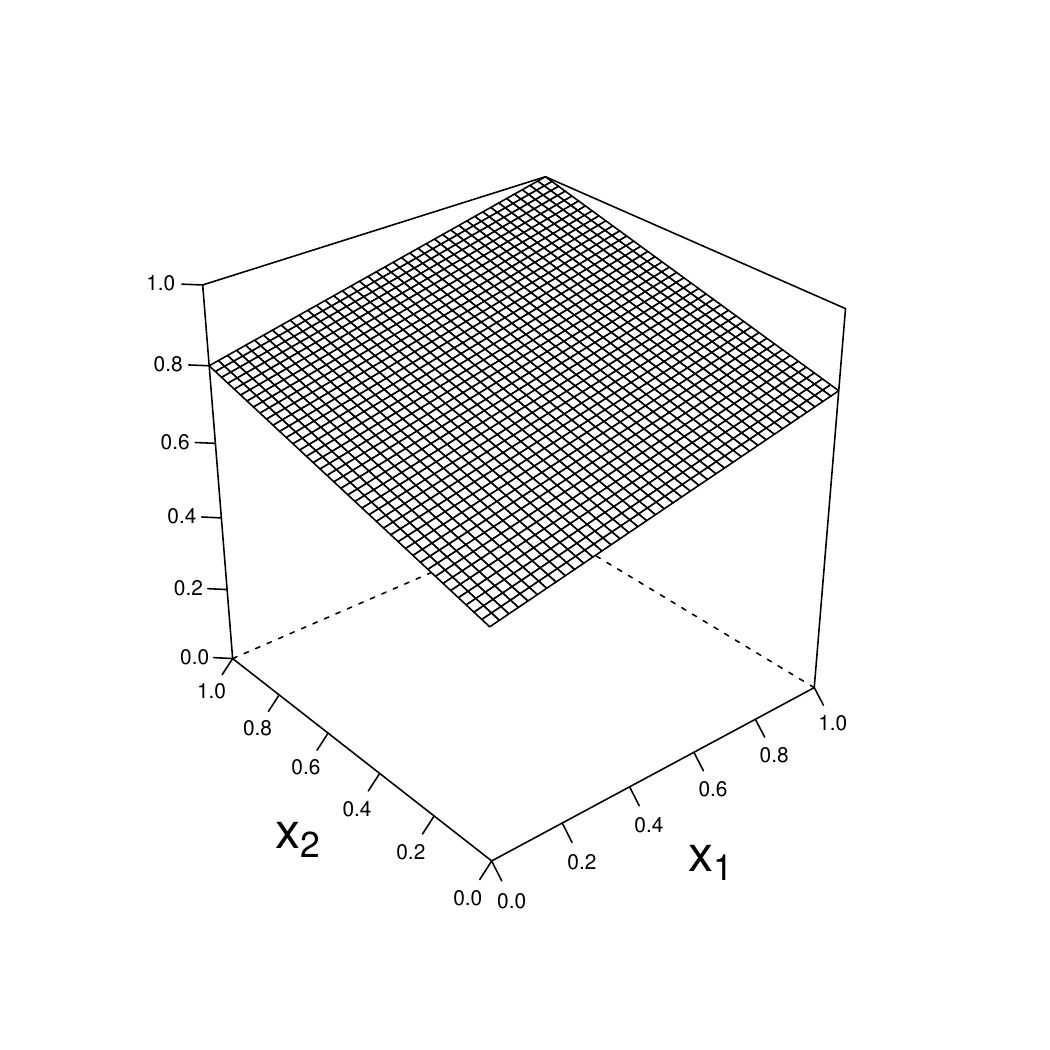}
\includegraphics[width=0.24\textwidth, trim={2cm 2cm 2cm 2cm}]{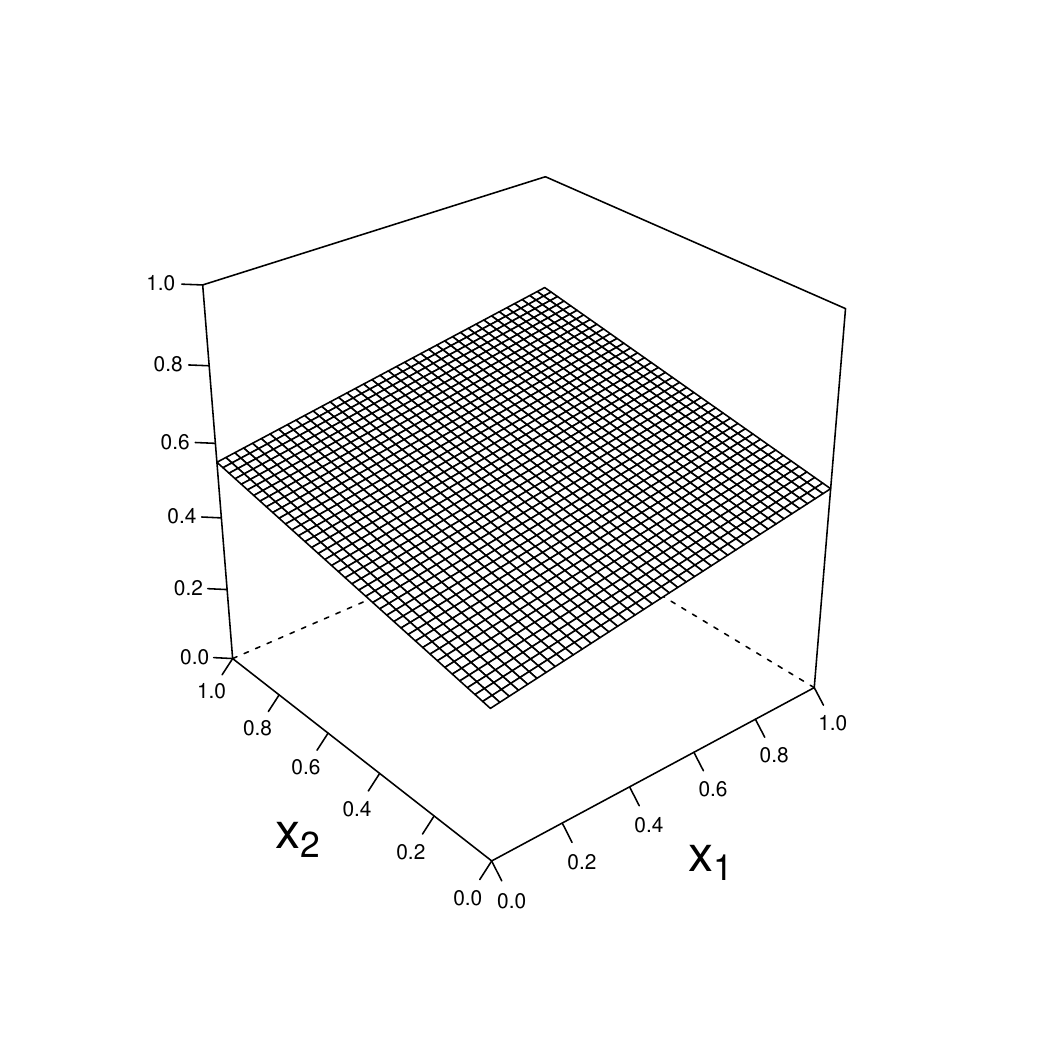}
\includegraphics[width=0.24\textwidth, trim={2cm 2cm 2cm 2cm}]{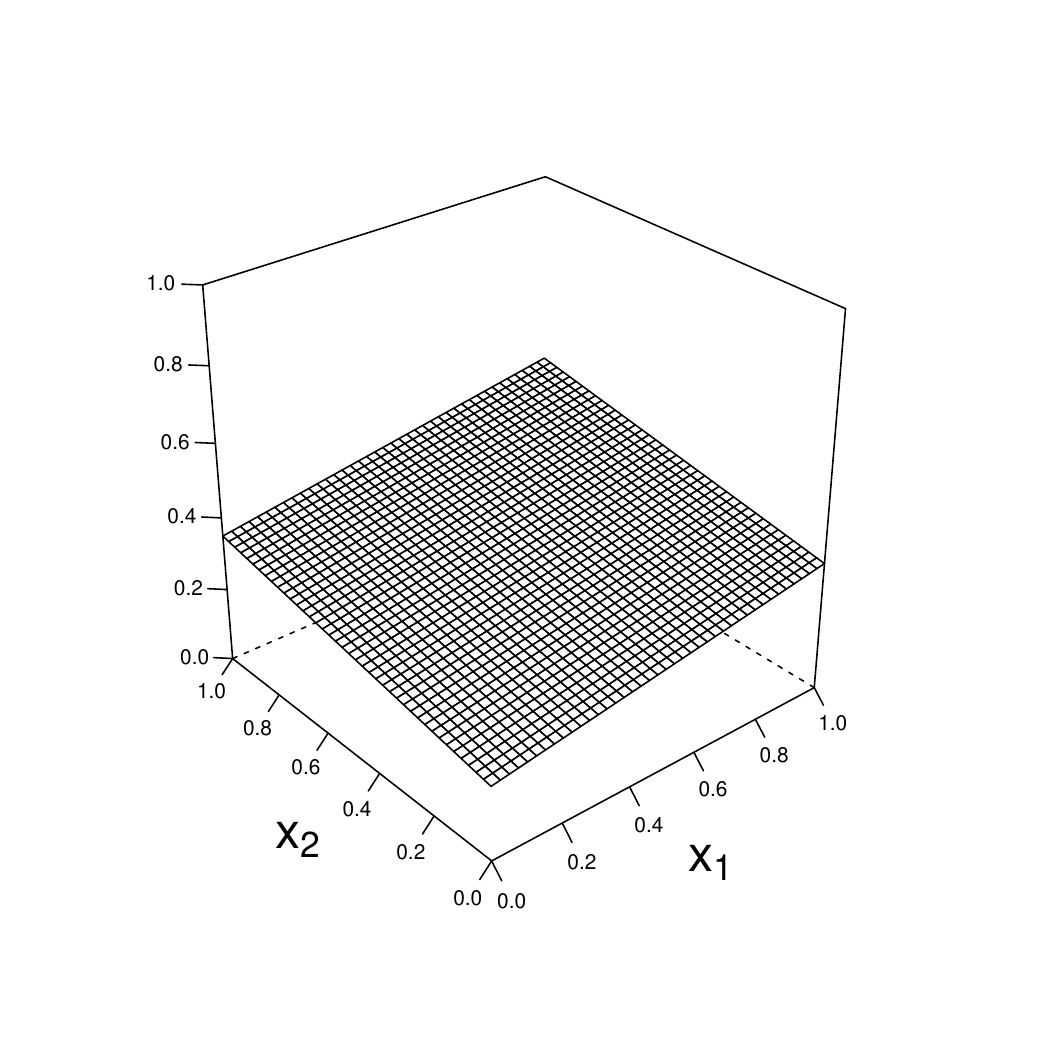}
\includegraphics[width=0.24\textwidth, trim={2cm 2cm 2cm 2cm}]{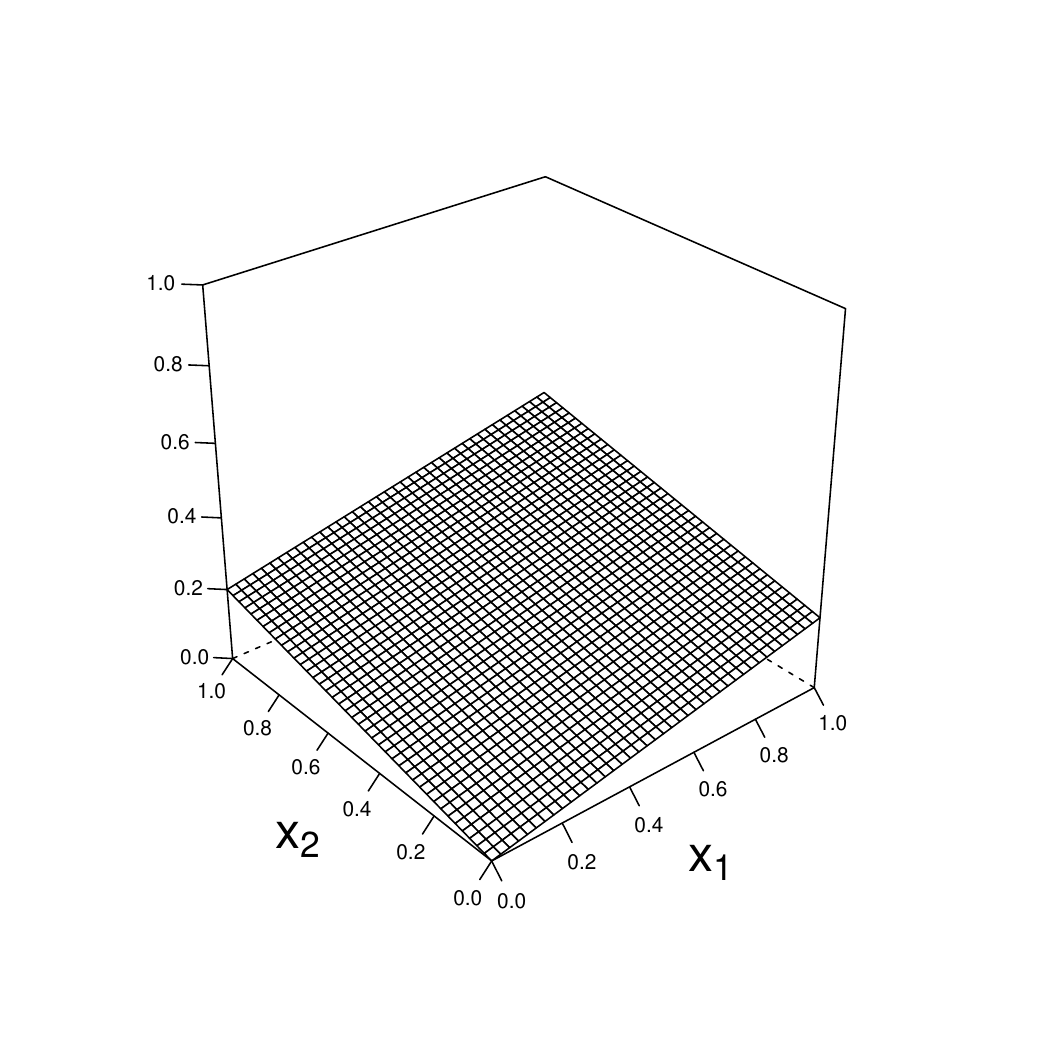}\\
\includegraphics[width=0.24\textwidth, trim={2cm 2cm 2cm 2cm}]{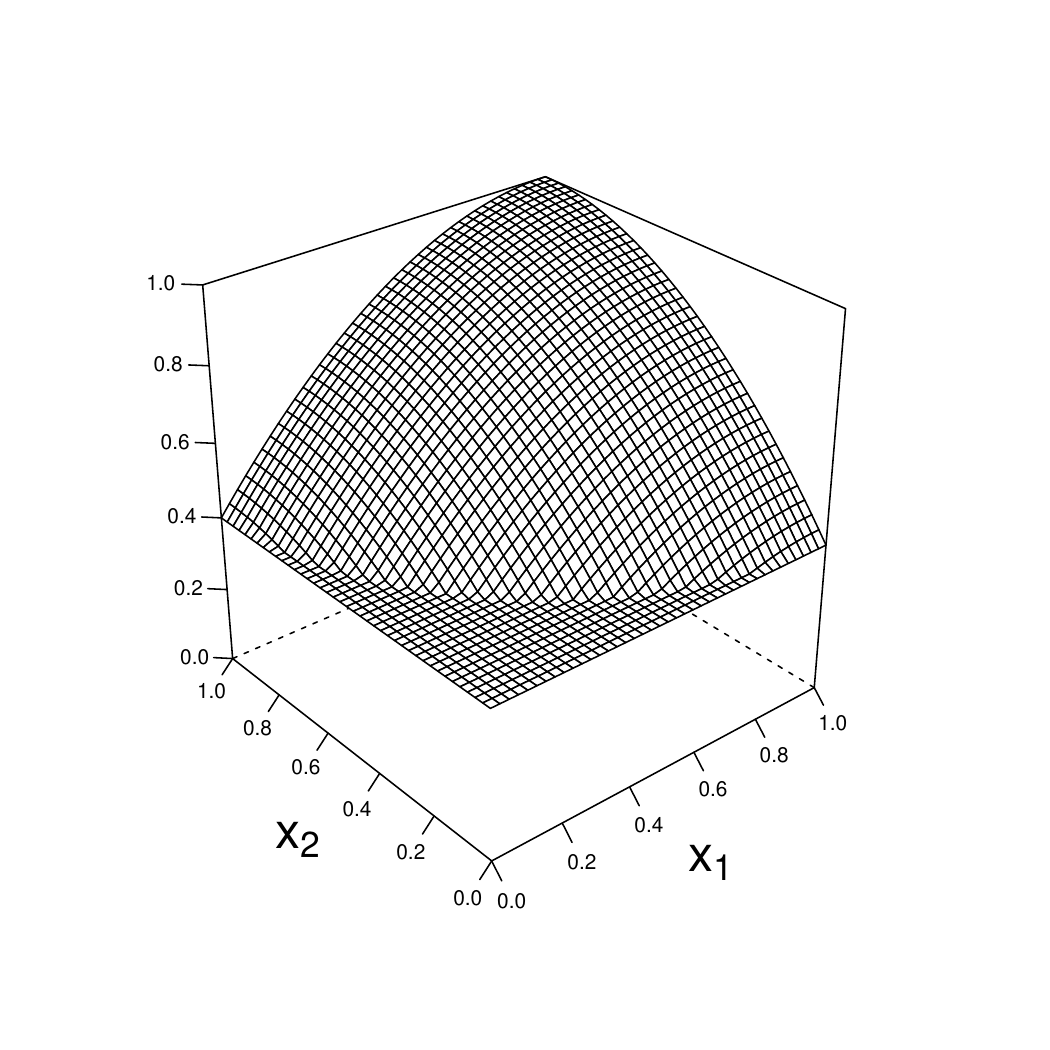}
\includegraphics[width=0.24\textwidth, trim={2cm 2cm 2cm 2cm}]{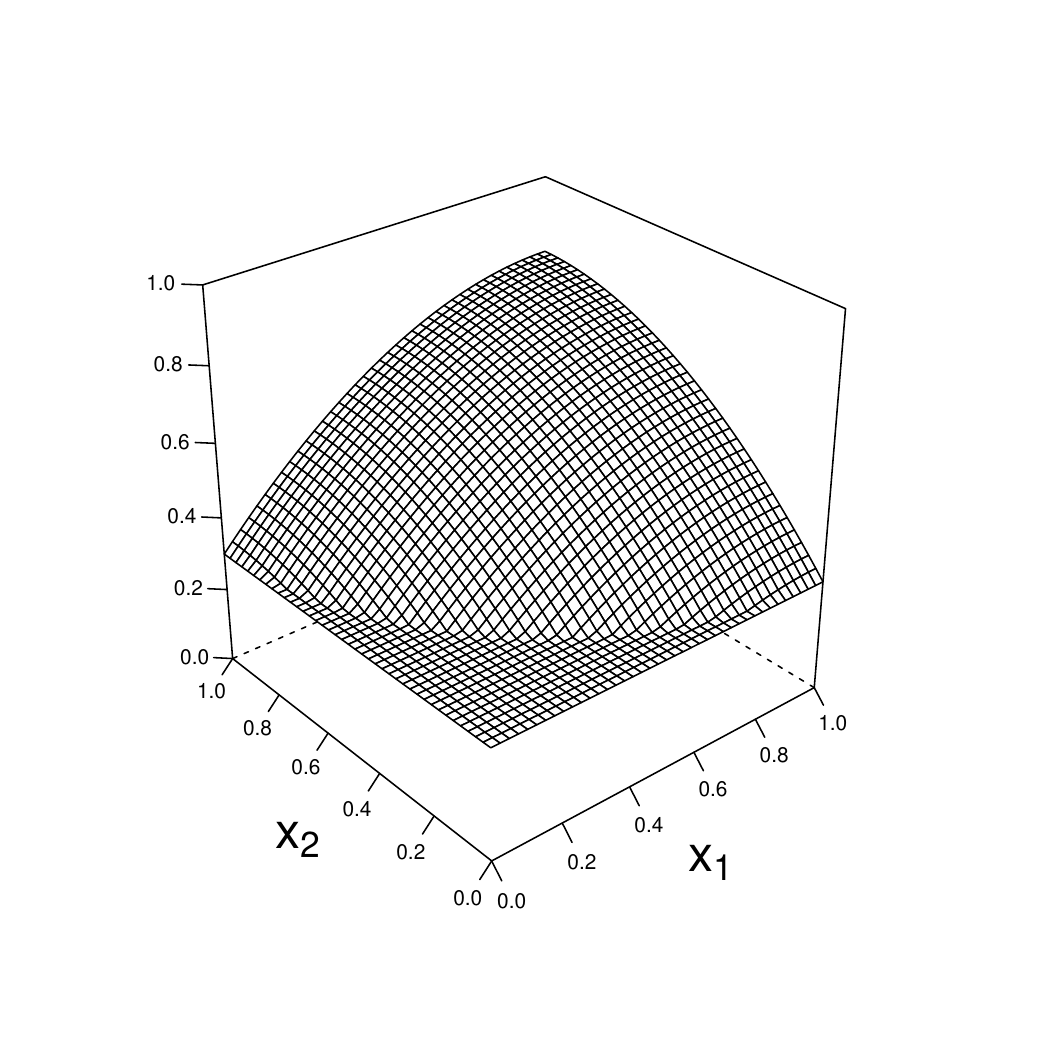}
\includegraphics[width=0.24\textwidth, trim={2cm 2cm 2cm 2cm}]{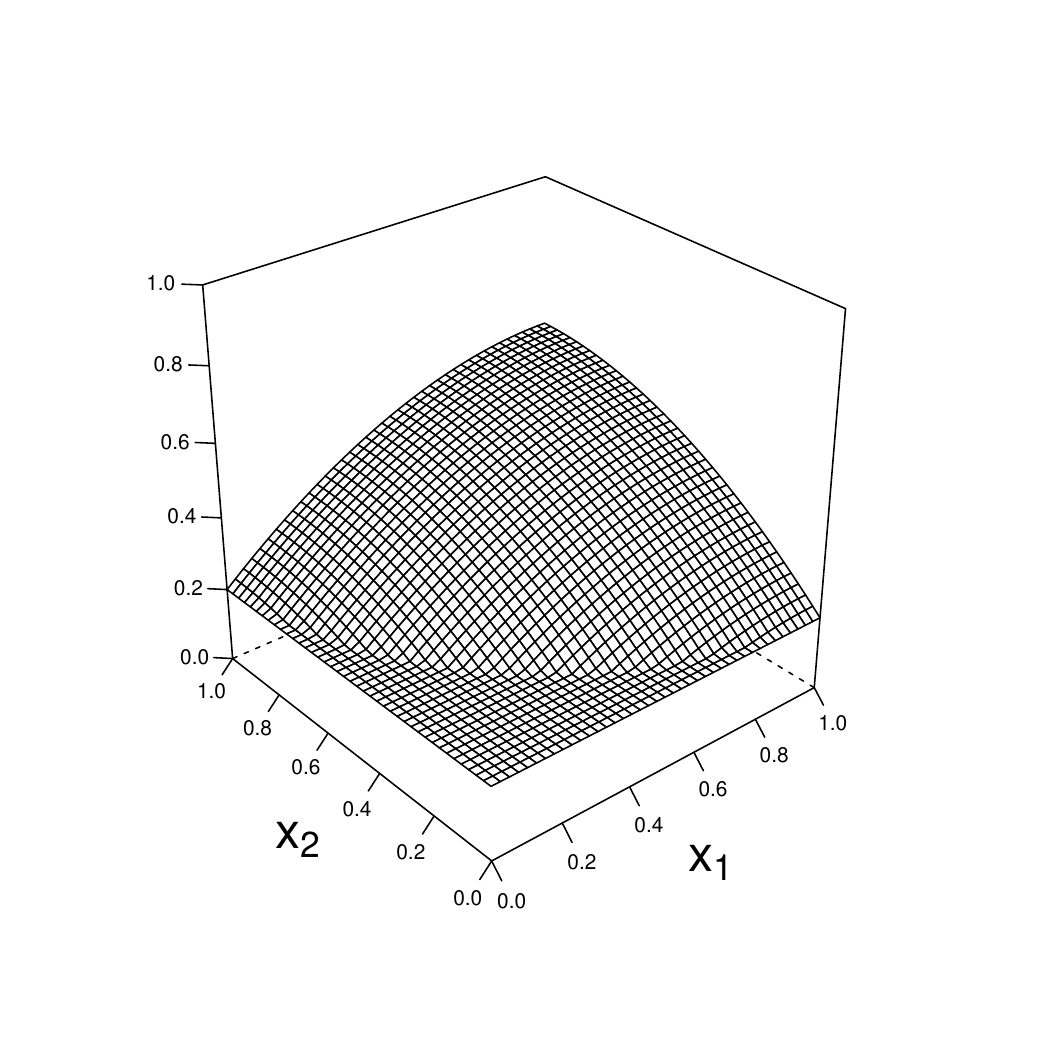}
\includegraphics[width=0.24\textwidth, trim={2cm 2cm 2cm 2cm}]{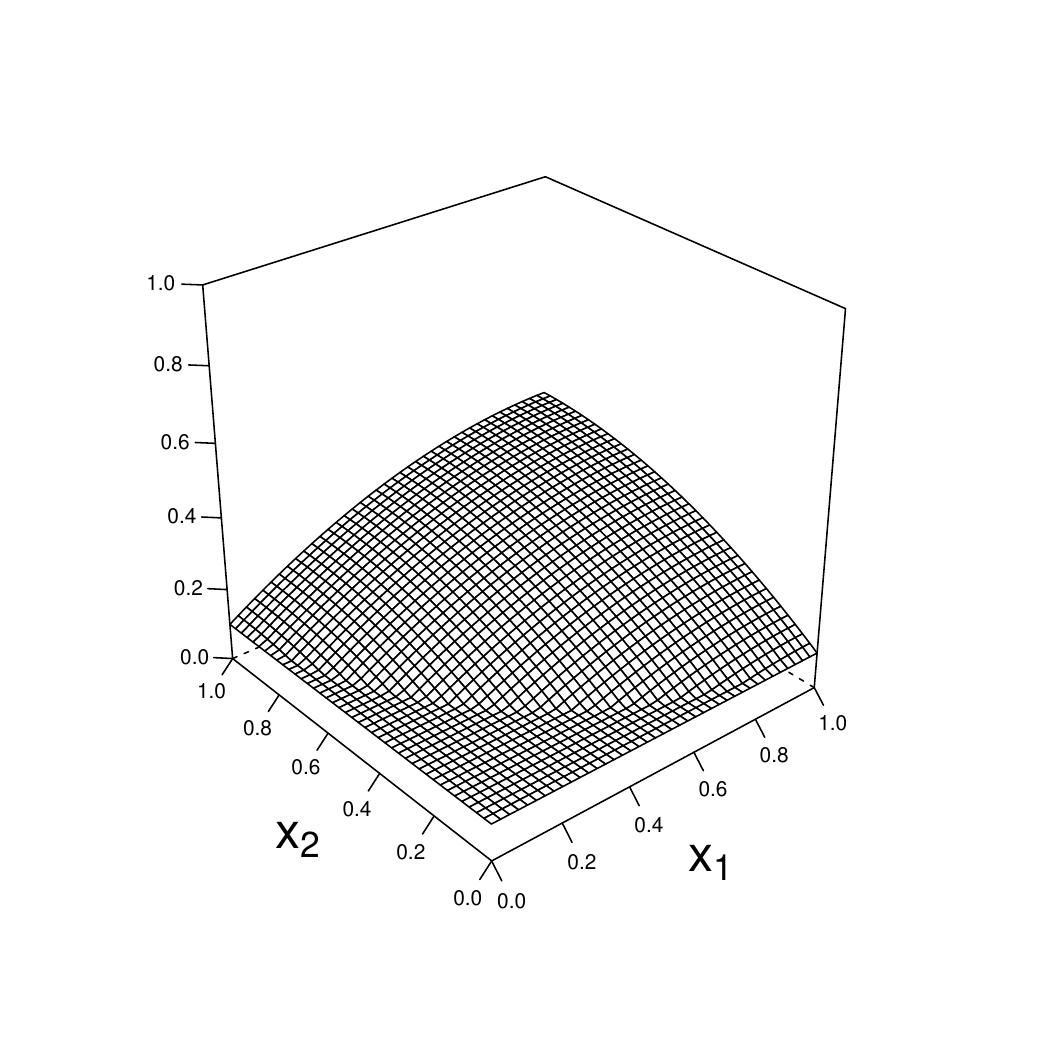}\\
\includegraphics[width=0.24\textwidth, trim={2cm 2cm 2cm 2cm}]{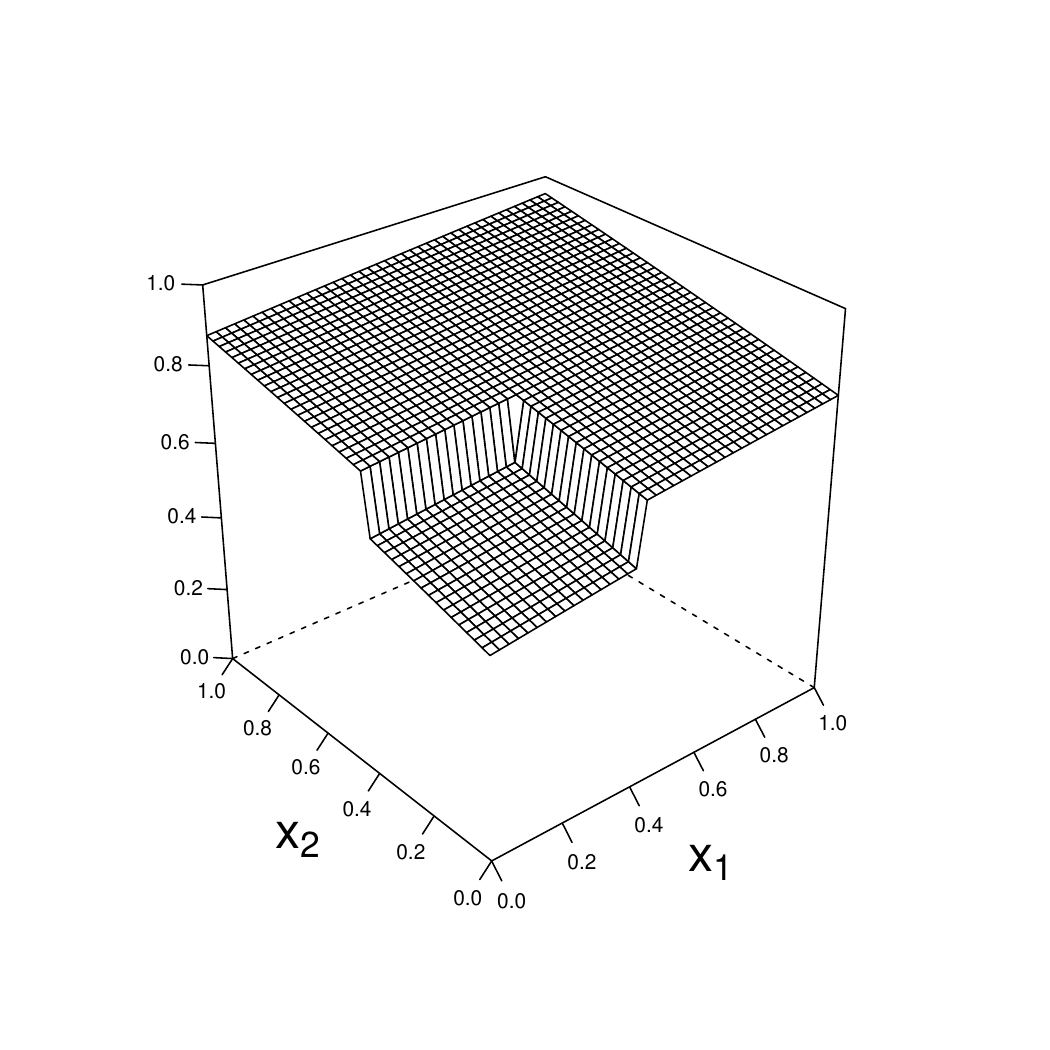}
\includegraphics[width=0.24\textwidth, trim={2cm 2cm 2cm 2cm}]{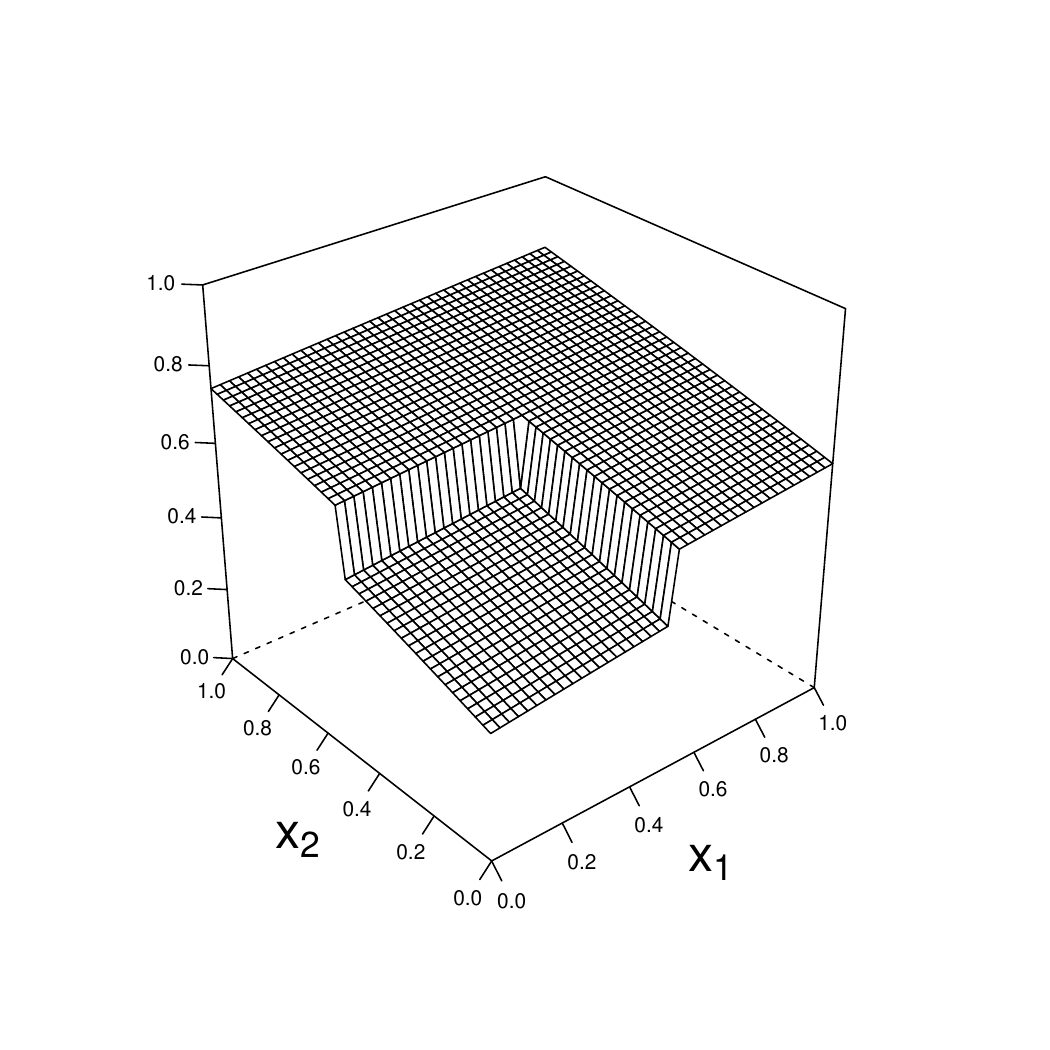}
\includegraphics[width=0.24\textwidth, trim={2cm 2cm 2cm 2cm}]{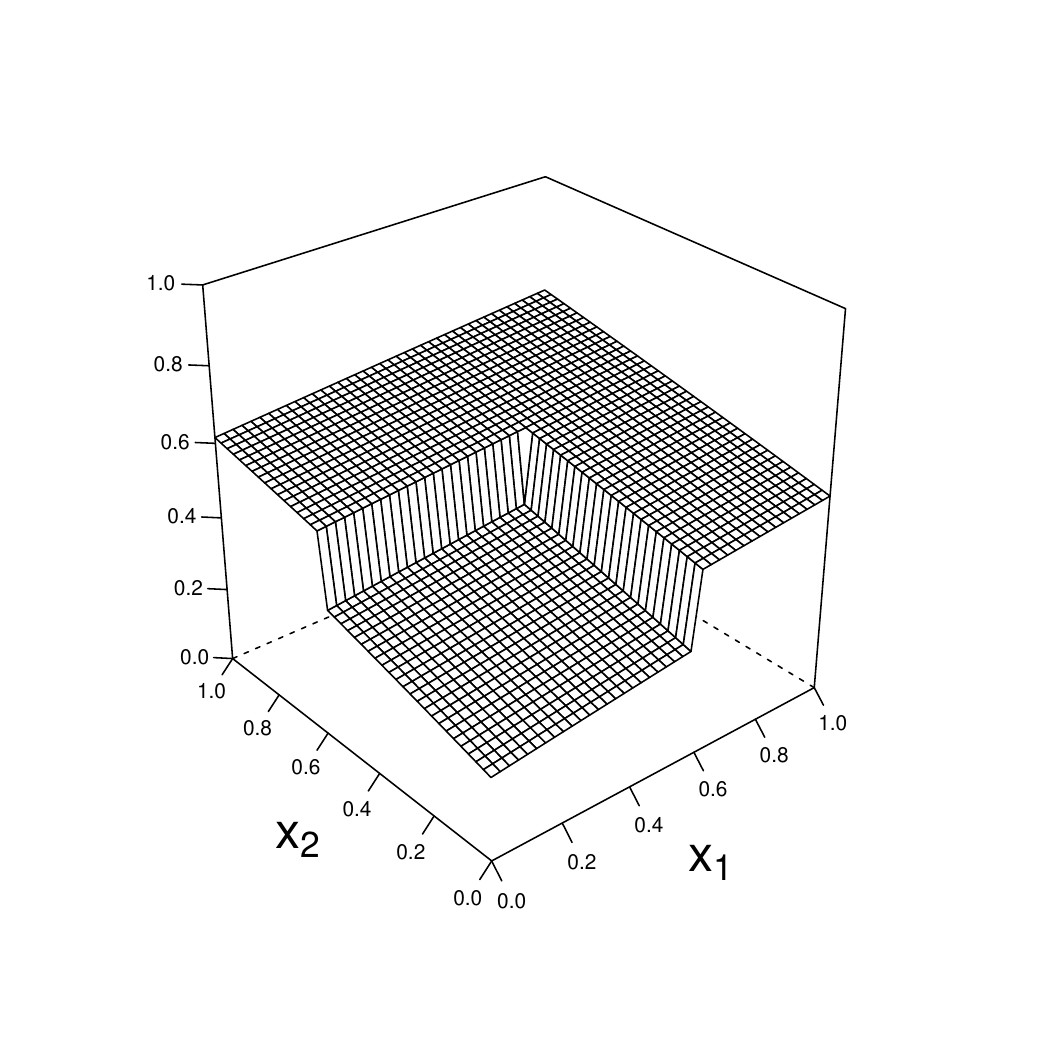}
\includegraphics[width=0.24\textwidth, trim={2cm 2cm 2cm 2cm}]{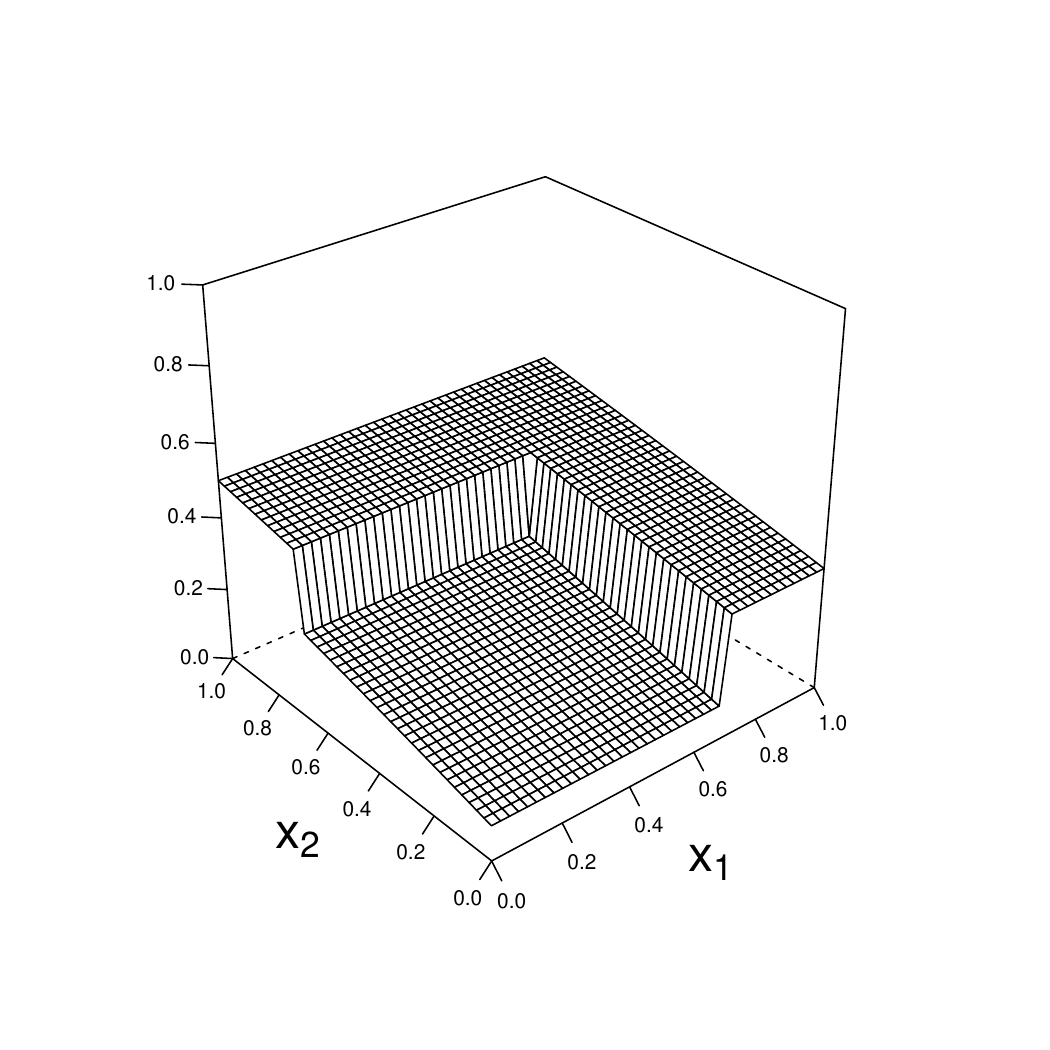}
\caption{Graphical displays of the survival functions $S(2\mid\cdot),\ldots,S(5\mid\cdot)$ (left to right) considered in Section~\ref{subsec:SimNonparametric}: linear functions (top), continuous  (middle), and discontinuous   (bottom).}
\label{fig:Sim31Truth}
\end{figure}

Figure~\ref{fig:Sim31Truth} shows three  sets of survival functions $S(2\mid\mathbf{x}), \ldots, S(5\mid\mathbf{x})$, which we use here as illustrations of the method; $S(1\mid\mathbf{x})=1$  and is therefore omitted from the figure. The exact definitions of these functions are provided in Section~S3.1 of the online supplementary material. In our first example, the functions $\lambda_2,\ldots,\lambda_5$ are linear, with $\lambda_2$ and $\lambda_5$ having identical slopes, and similarly for $\lambda_3$ and $\lambda_4$. In the second example, the survival functions are constant when one of two predictors has a small value, and increase continuously otherwise. In the third example the survival functions are discontinuous, with each function $S(2\mid\mathbf{x}),\ldots , S(5\mid\mathbf{x})$ having a different set of discontinuity points. The proportions of data points in the different categories varied: in the first example, there were approximately similar numbers of data points in each category; in the second, the largest number of points were in category 1, followed by category 5; while in the third example, category 5 had the largest number of data points. 

\begin{figure}
    \centering
    \includegraphics[width=0.24\textwidth, trim={2cm 2cm 2cm 2cm}]{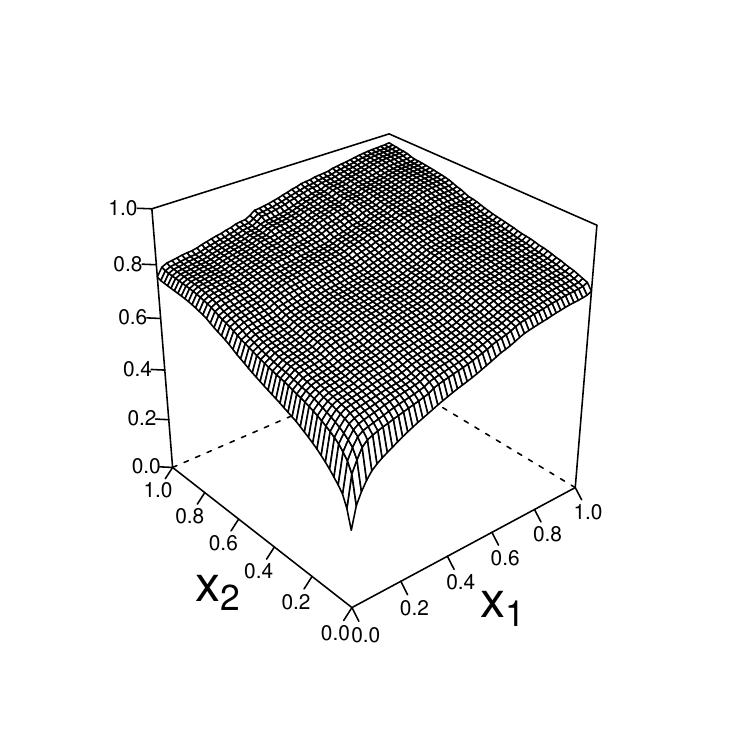}
    \includegraphics[width=0.24\textwidth, trim={2cm 2cm 2cm 2cm}]{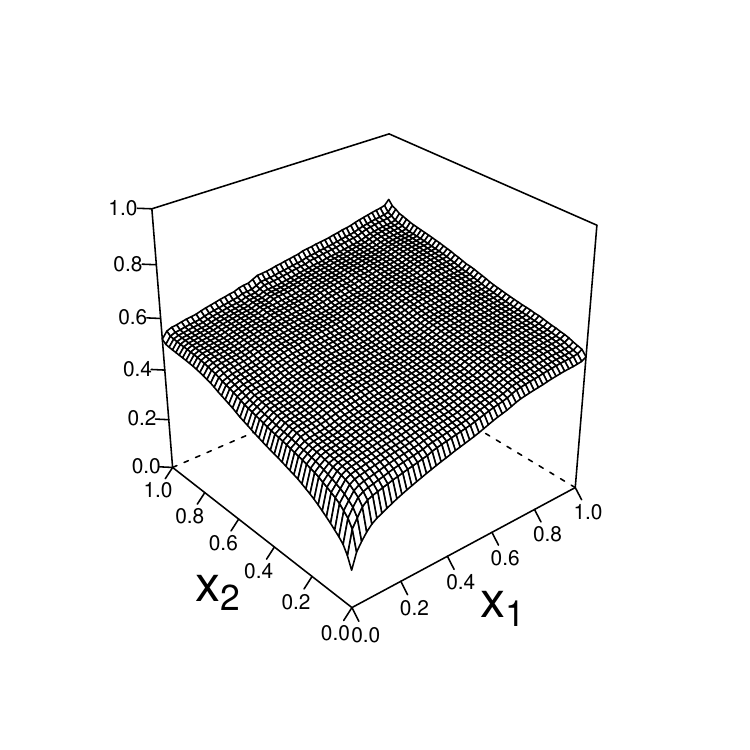}
    \includegraphics[width=0.24\textwidth, trim={2cm 2cm 2cm 2cm}]{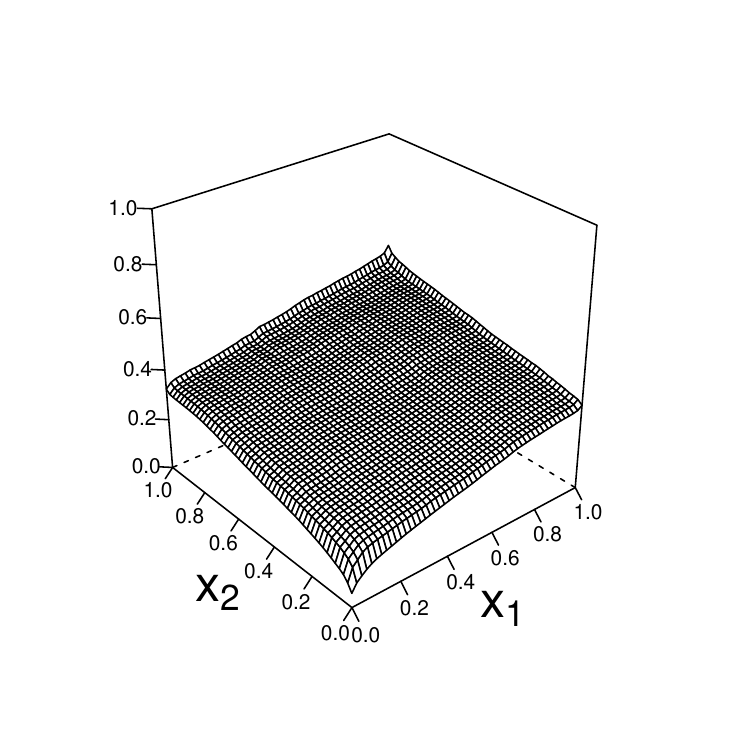}
    \includegraphics[width=0.24\textwidth, trim={2cm 2cm 2cm 2cm}]{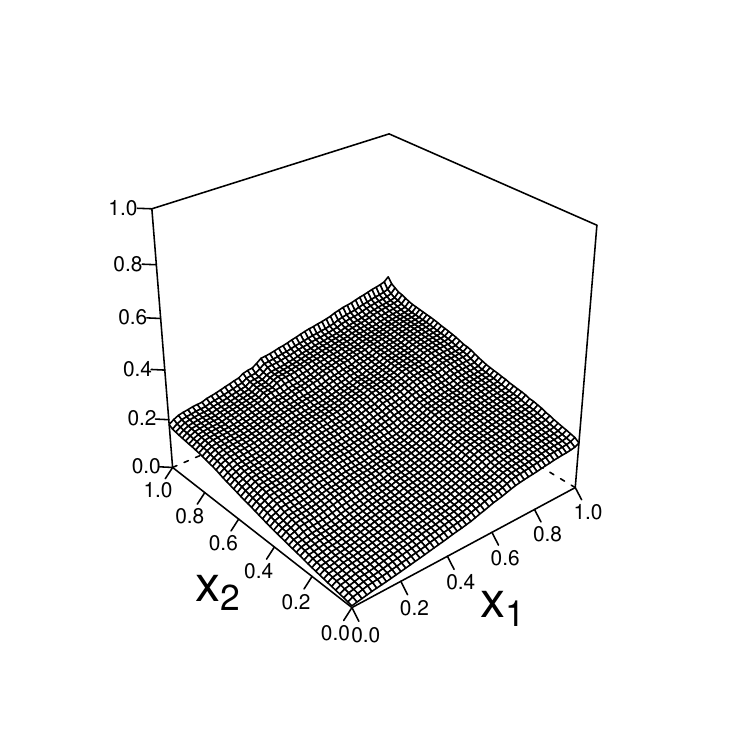}\\[5pt]
    \includegraphics[width=0.24\textwidth, trim={2cm 2cm 2cm 2cm}]{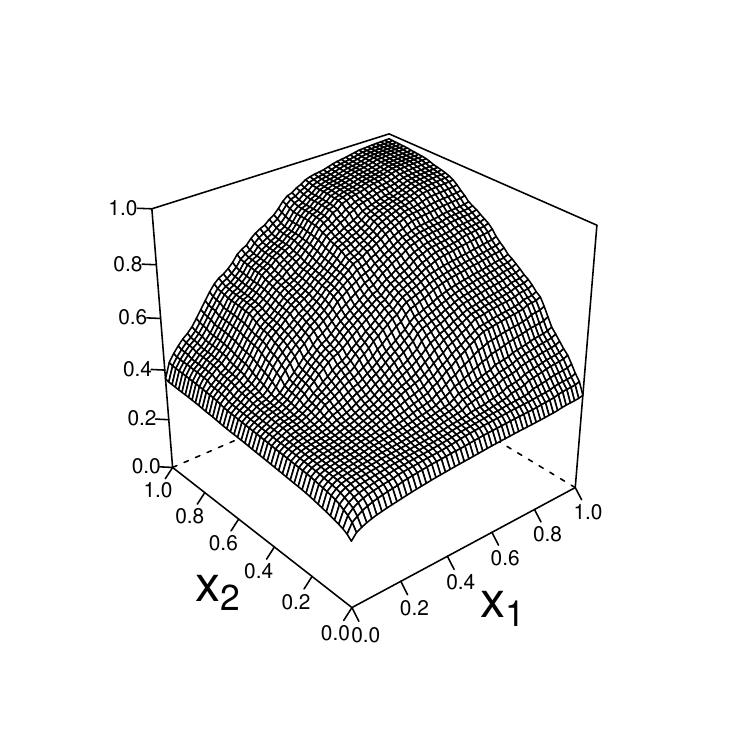}
    \includegraphics[width=0.24\textwidth, trim={2cm 2cm 2cm 2cm}]{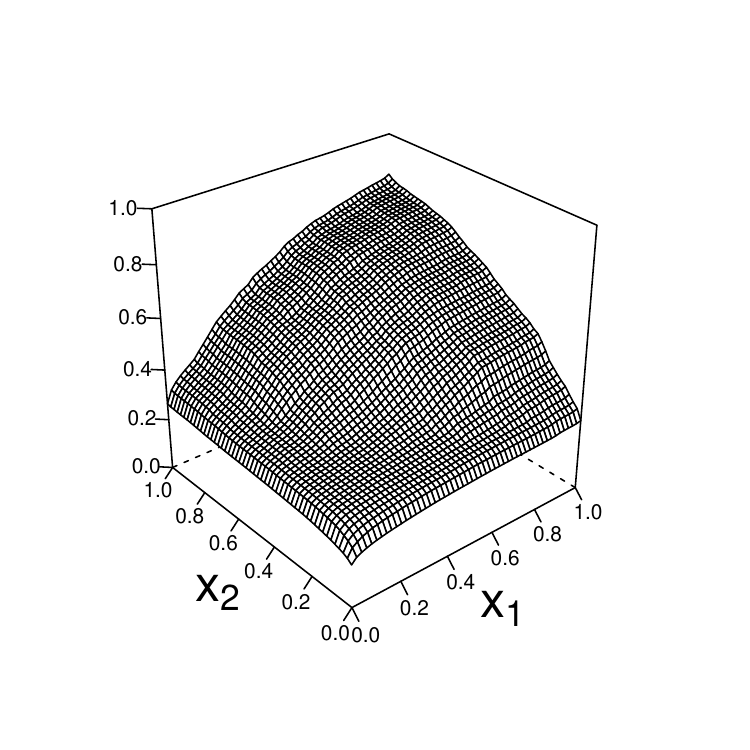}
    \includegraphics[width=0.24\textwidth, trim={2cm 2cm 2cm 2cm}]{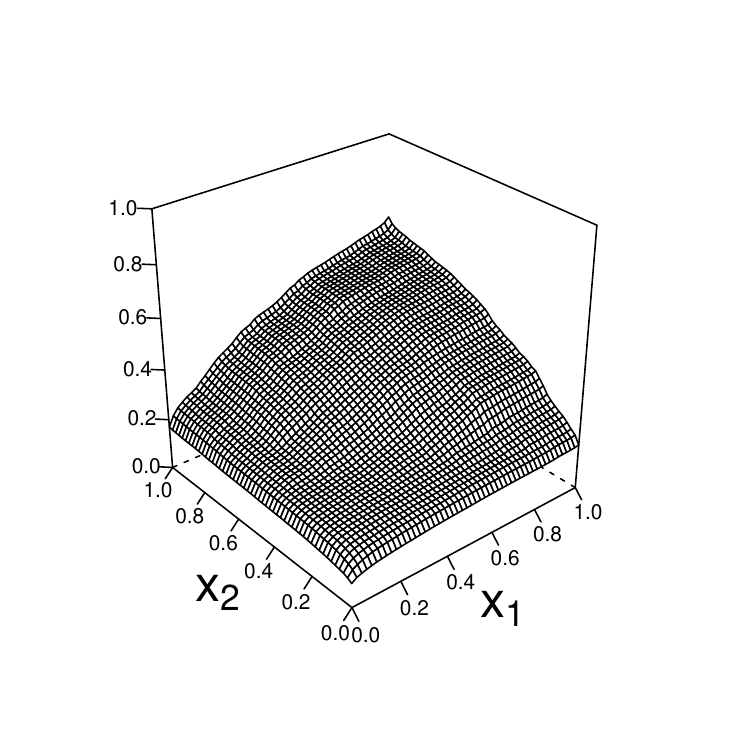}
    \includegraphics[width=0.24\textwidth, trim={2cm 2cm 2cm 2cm}]{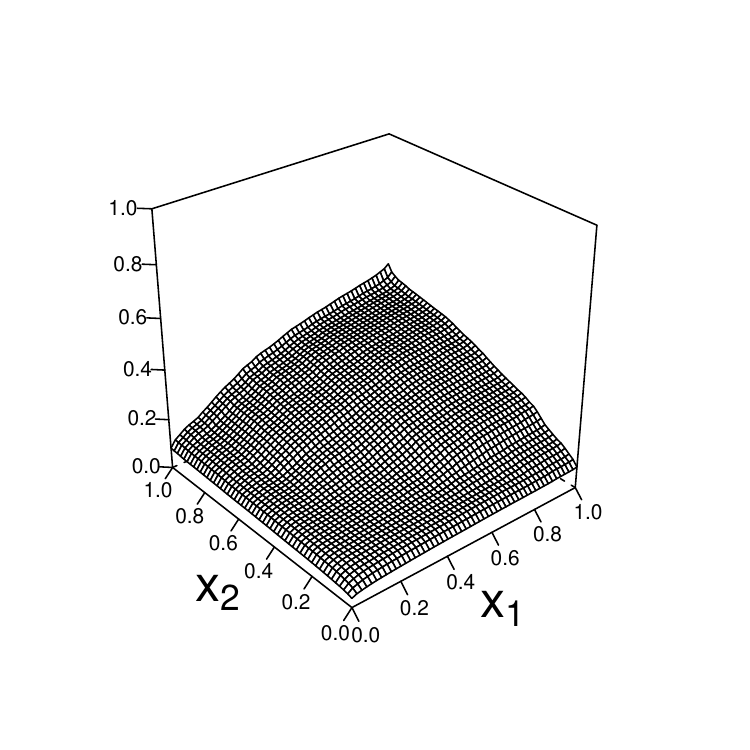}\\[5pt]
    \includegraphics[width=0.24\textwidth, trim={2cm 2cm 2cm 2cm}]{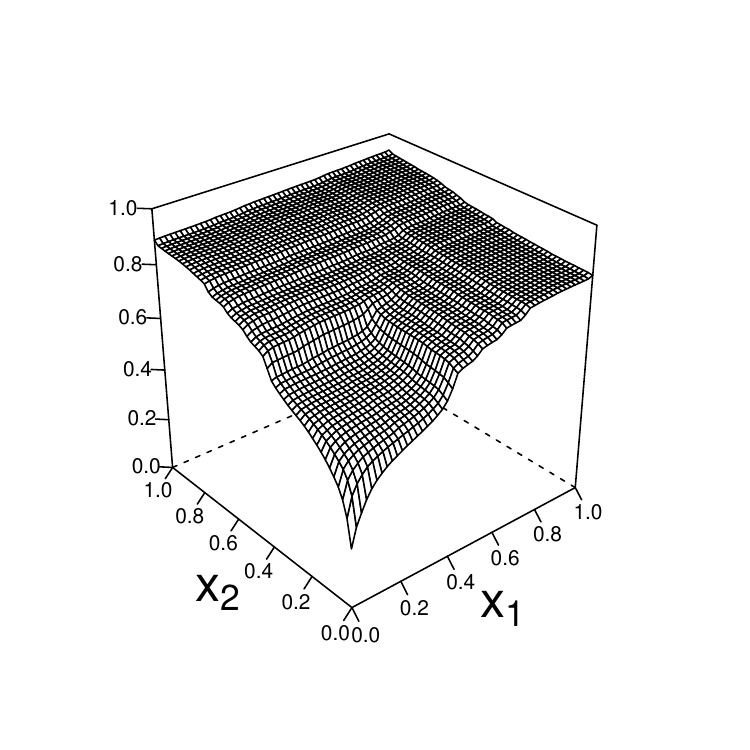}
    \includegraphics[width=0.24\textwidth, trim={2cm 2cm 2cm 2cm}]{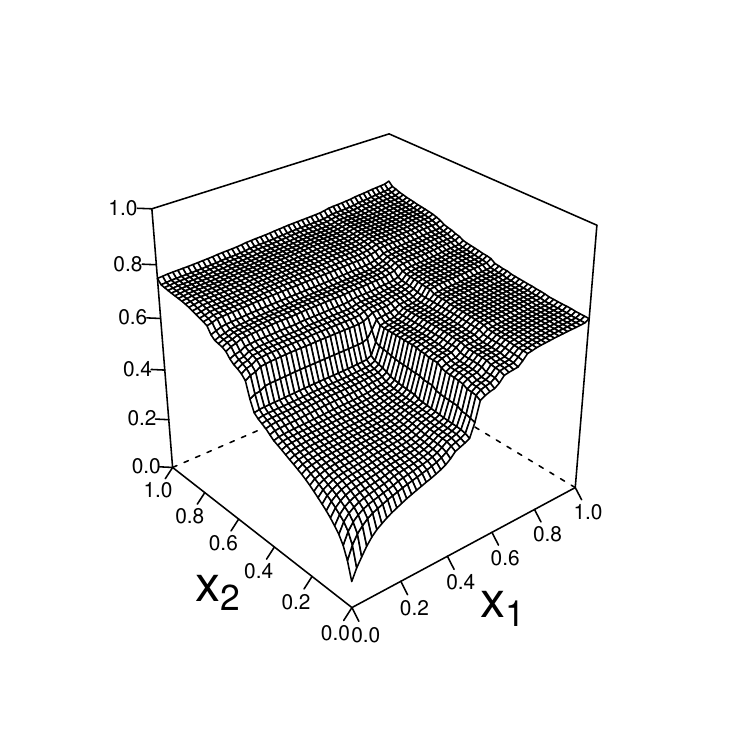}
    \includegraphics[width=0.24\textwidth, trim={2cm 2cm 2cm 2cm}]{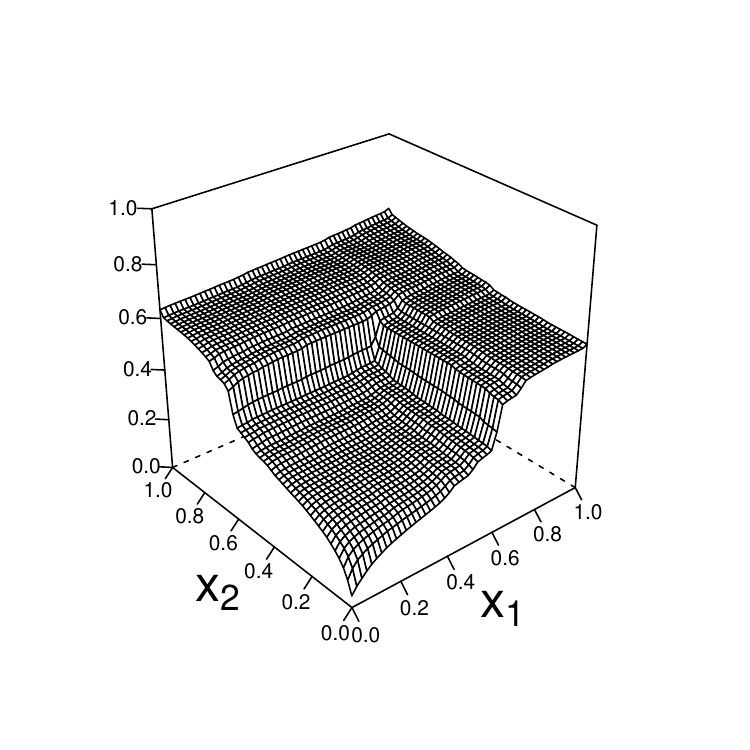}
    \includegraphics[width=0.24\textwidth, trim={2cm 2cm 2cm 2cm}]{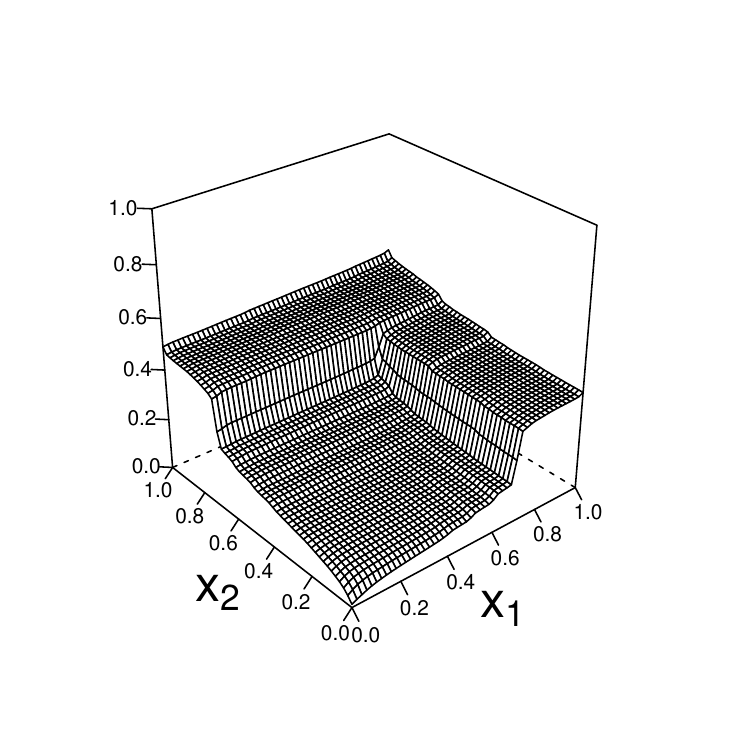}\\
    \caption{Graphical displays of the posterior mean estimates of  the survival functions $S(2\mid\cdot),\ldots,S(5\mid\cdot)$ in Section~\ref{subsec:SimNonparametric}, averaged over results from 20 simulated data sets of size $N=5000$: linear (top), continuous (middle), and discontinuous (bottom).}
    \label{fig:Sim1PosteriorEstimates}
\end{figure}

\begin{table}
\centering
\caption{Error measures MAE($k$)$\times10^2$ and MAE$\times 10^2$, with standard deviations$\times10^2$, computed from 20 sets of simulated data from the models in Figure~\ref{fig:Sim31Truth}, for $n = 1000$ and $n = 5000$.}
\vspace{0.2cm}
\label{tab:Sim31Results}
\begin{tabular}{l c | c c c c c c}
\hline
Functions& N & MAE(1) & MAE(2) & MAE(3) & MAE(4) & MAE(5) & MAE\\[2pt]
\hline
Linear&1000&3.8\,(1.1)&2.1\,(0.6)&1.8\,(0.5)&2.2\,(0.6)&3.0\,(0.7)&4.1\,(0.2)\\ [2pt]
      &5000&2.2\,(0.4)&1.3\,(0.3)&1.3\,(0.3)&1.4\,(0.3)&2.0\,(0.3)&2.6\,(0.1)\\
[4pt]
Cont. &1000&4.3\,(0.8)&2.9\,(0.7)&2.3\,(0.6)&2.1\,(0.6)&3.0\,(0.8)&4.6\,(0.2)\\
[2pt]
      &5000&2.9\,(0.4)&1.6\,(0.3)&1.5\,(0.4)&1.4\,(0.4)&1.8\,(0.4)&2.9\,(0.2)\\ 
[4pt]
Discont.&1000&4.4\,(0.9)&3.7\,(0.4)&4.2\,(0.6)&4.8\,(0.5)&3.7\,(0.7) &4.7\,(0.3)\\[2pt]
    &5000&2.7\,(0.4)&2.2\,(0.3)&2.1\,(0.4)&2.5\,(0.3)&2.1\,(0.4)& 2.7\,(0.2)\\
    [2pt]
\hline
\end{tabular}
\end{table}

\begin{figure}
\centering
\includegraphics[width=0.4\textwidth, trim={1cm 1cm 0cm 0cm}] {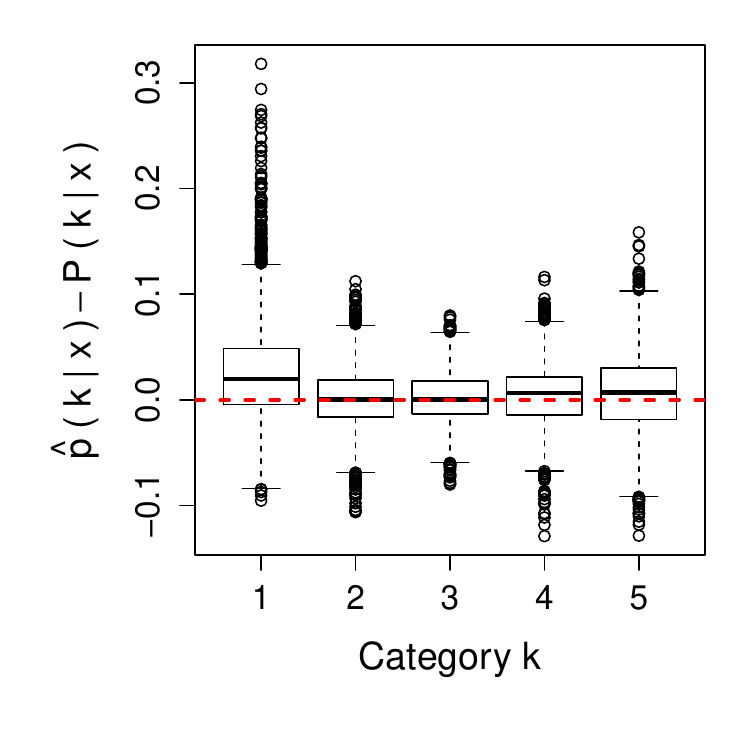}
\hspace{1cm}
\includegraphics[width=0.4\textwidth, trim={1cm 1cm 0cm 0cm}] {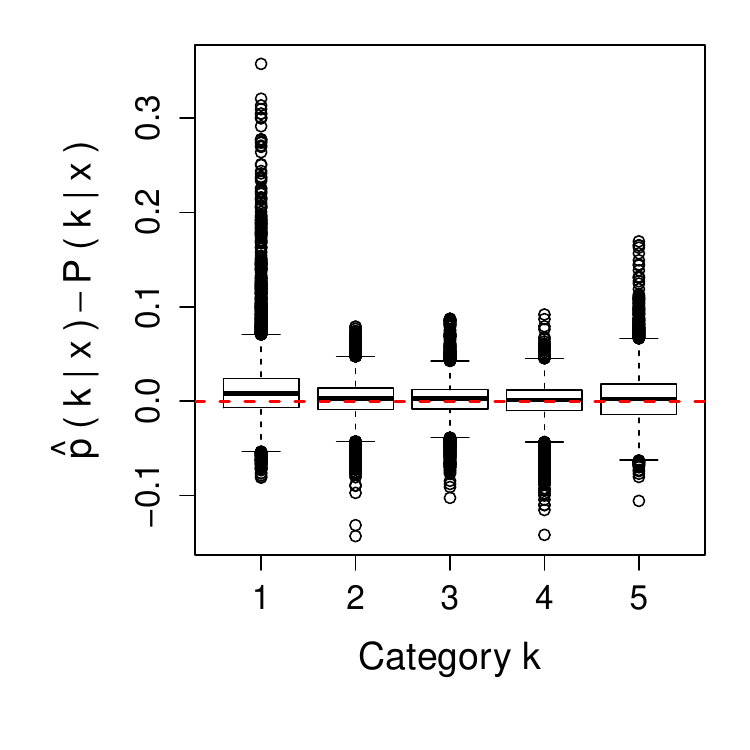}\\
\includegraphics[width=0.4\textwidth, trim={1cm 1cm 0cm 0cm}] {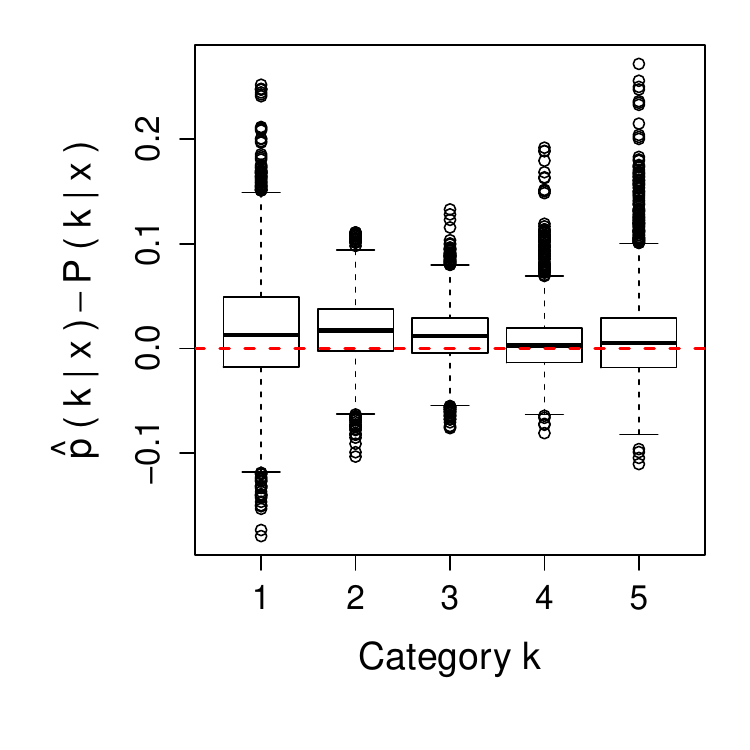}
\hspace{1cm}
\includegraphics[width=0.4\textwidth, trim={1cm 1cm 0cm 0cm}] {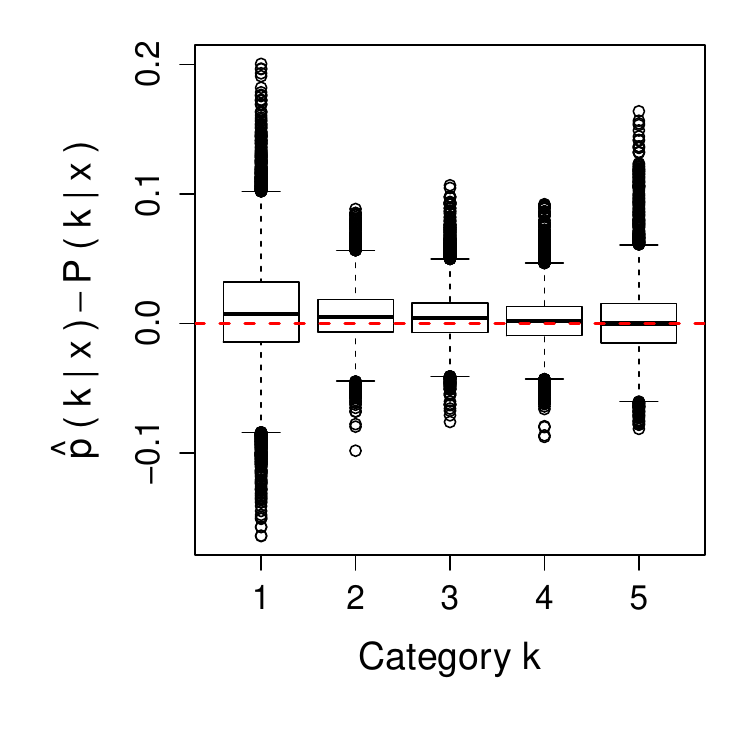}\\
\includegraphics[width=0.4\textwidth, trim={1cm 1cm 0cm 0cm}] {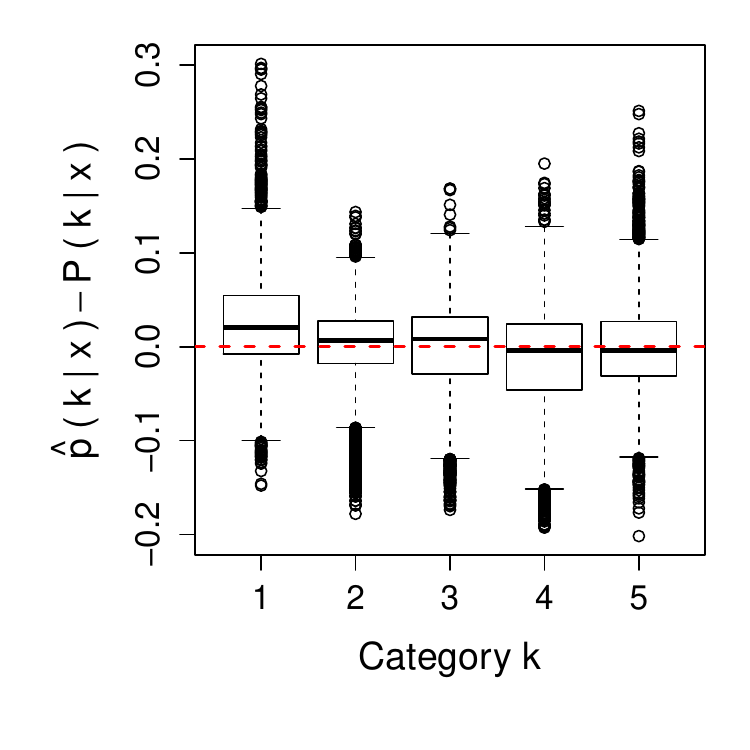}
\hspace{1cm}
\includegraphics[width=0.4\textwidth, trim={1cm 1cm 0cm 0cm}] {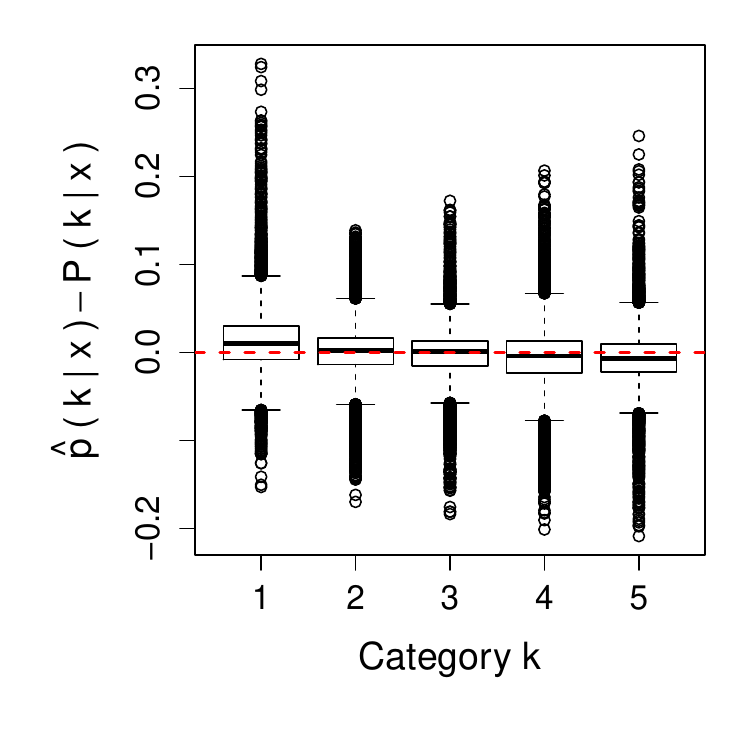}
\caption{Boxplots of the difference $\hat{p}(Y=y_n\mid\mathbf{x}_n) - P(Y=y_n\mid\mathbf{x}_n)$ in the estimated and true probabilities for the $K=5$ categories across 20 simulated data sets with $N=1000$~(left) and $N=5000$~(right) data points for the survival functions in Figure~\ref{fig:Sim31Truth} (top to bottom).}
\label{fig:Sim1Boxplots}
\end{figure}

The posterior estimates in Figure~\ref{fig:Sim1PosteriorEstimates} show that our methodology correctly identifies the overall structure of the underlying model functions in Figure~\ref{fig:Sim31Truth}; more plots illustrating the uncertainty of the model estimates and general model fit are provided in Section~S2.3 of the online Supplementary Material. From Table~\ref{tab:Sim31Results} we can see that the considered error measures MAE($k$) and MAE, and their variances, became smaller when the size of the data sets increased from 1000 to 5000. This, together with many additional numerical experiments we carried out, provides some empirical evidence that the model and the estimation algorithm are working correctly in the sense of $\hat{p}(k\mid\mathbf{x}_n)$ providing consistent estimates of the true values $P(k\mid\mathbf{x}_n)$ for large data sets. Moreover, Figure~\ref{fig:Sim1Boxplots} shows that the differences between the estimated and true probabilities are approximately symmetrically distributed around 0 in the middle categories $2, 3$ and $4$; however, particularly in category 1 there appears to be some bias towards positive values. A closer examination of the covariate values with the largest differences between true and estimated values reveals that many of them are located close to the boundaries of the unit square, in particular, close to the origin $\mathbf{x}=(0,0)$; see Supplementary Figure~S3 for the data points with the largest estimation errors.

The poor fit very close to the boundaries can be explained as follows. Suppose that $(y_1,\mathbf{x}_1)$ is the observation with the smallest value for the first covariate $x_1$. Then, the estimate maximizing the likelihood value for this observation would be $\hat\lambda_k(\mathbf{x}_1)=0$ for $k>y_1$ and $\hat\lambda_k(\mathbf{x}_1)=1$ for $k\leq y_1$. In case $y_1=1$, the likelihood function thus pushes the estimates towards $\hat\lambda_k(\mathbf{x}_1)=0$ for all $k$, and as $\mathbf{x}_1$ has the smallest value for $x_1$, there are no other data from which to draw inference. Consequently, the estimated functions will be close to 0 if $y_1=1$. If $y_1\neq 1$, this is less problematic, because the effect of the data point $(y_1,\mathbf{x}_1)$ favouring $\hat{\lambda}_2 = 1$ would be mitigated by the other data points. A similar argument can be made for values close to $x_1=1$ or $x_2=1$ and $y_t=K$. These difficulties are ameliorated to some extent when using the conditional prior proposed in Section~\ref{section:spiking} with the user-specified parameter $d$ set to $d=K=5$. A comparison of the posterior estimates for $\lambda_1,\ldots,\lambda_5$ is provided in Section~S3.4 of the online Supplementary Material.
 
\subsection{Semi-parametric model structures}
\label{subsec:SimSemiparametric}

As in the non-parametric case, we consider three sets of functions for the semi-parametric model structure. The monotonic functions $\lambda_2,\ldots,\lambda_5$ have the same shape as in Figure~\ref{fig:Sim31Truth}, but with the functional levels scaled to the interval $[-2,2]$ instead of $[0,1]$. These functions result in categories 1 and 5 being observed the most in all three studies; the proportions of observations in category 1 are 0.25, 0.39 and 0.25 for the set of linear, continuous and discontinuous functions respectively, while these proportions are 0.25, 0.27 and 0.34 for category 5. The vector $\bm\beta$ of linear regression coefficients in expression~\eqref{eq:ModelSemiparametric} is fixed to $\bm\beta=(\beta_1,\beta_2,\beta_3) = (0.3,-0.5,0.1)$ and the covariates are independently and standard normally distributed, $\mathbf{z}\sim\mbox{MVN}\left(0, I_{3\times3}\right)$, across all studies. This setup yields that the survival functions have shapes similar to those in Section~\ref{subsec:SimNonparametric}, with the covariates $\mathbf{z}$ having a moderate effect on the probabilities for the different categories. Plots of the survival functions and an illustration of the effect of the values in $\mathbf{z}$ are provided in Section~S4.1 in the online Supplementary Material.

\begin{table}
\centering
\caption{Error measures MAE($k$)$\times10^2$ and MAE$\times 10^2$, with standard deviations$\times10^2$, computed from 20 sets of simulated data for the semi-parametric models in Section~\ref{subsec:SimSemiparametric}, for $N = 1000$ and $N = 5000$ data points.}
\vspace{0.2cm}
\label{tab:Sim32Results}
\begin{tabular}{l c | c c c c c c}
\hline
Functions& N & MAE(1) & MAE(2) & MAE(3) & MAE(4) & MAE(5) & MAE\\[2pt]
\hline
Linear&1000&3.7\,(0.8)&2.0\,(0.6)&1.8\,(0.7)&1.8\,(0.7)&3.3\,(0.6)&3.6\,(0.3)\\[2pt]
      &5000&2.1\,(0.4)&1.2\,(0.4)&1.1\,(0.3)&1.0\,(0.3)&2.2\,(0.4)&2.2\,(0.1)\\[4pt]
Cont. &1000&4.7\,(0.7)&2.3\,(0.5)&2.0\,(0.5)&2.3\,(0.9)&3.1\,(0.7)&4.2\,(0.3)\\[2pt]
      &5000&3.2\,(0.5)&1.5\,(0.3)&1.4\,(0.3)&1.4\,(0.3)&2.0\,(0.4)&2.6\,(0.1)\\[4pt] 
Discont.&1000&4.2\,(0.6)&3.8\,(0.6)&4.2\,(0.5)&4.7\,(0.6)&3.9\,(0.8)&4.6\,(0.3)\\[2pt]
        &5000&2.7\,(0.5)&2.3\,(0.3)&2.1\,(0.3)&2.4\,(0.3)&2.3\,(0.4)&2.5\,(0.2)\\[2pt] 
\hline
\end{tabular}
\end{table}

Table~\ref{tab:Sim32Results} shows again that the considered error measures MAE($k$) and MAE, and their variances, are decreasing with an increasing sample size. Moreover, the error measures are similar to the ones in Table~\ref{tab:Sim31Results}. By plotting the posterior mean estimates, we find that the function estimates $\hat{\lambda}_2,\ldots,\hat\lambda_5$ strongly resemble the true functions, and that the 95\% central credibility intervals for $\bm\beta$ include the true values in all but one repetition; posterior density plots for $\bm\beta$ are shown in Supplementary Figure~S7. As for the non-parametric case, the highest differences between the estimated and true probabilities, $\hat{p}(y_n\mid\mathbf{x}_n)$ and $P(y_n\mid \mathbf{x}_n)$~($n=1,\ldots,N$), occur for the first and last category; box plots of the difference are provided in Supplementary Figure~S8. Based on this, our conclusion is that the proposed approach works well for both non-parametric and semi-parametric model structures.

\section{Illustrations with real data}\label{sec:data}

\subsection{PISA schools data set}\label{sec:pisa}

With our first real data example, we wanted to illustrate (i) the ability of the proposed approach to accommodate both continuous and ordered categorical covariates, (ii) the use of semi-parametric formulations allowing for cluster-level random effects as outlined in Section \ref{sec:SemiparametricModel}, and (iii) the graphical presentation of the model output in 3-dimensional surface plots. For this purpose, we analyzed data from the 2015 PISA school questionnaire \citep{oecd2016pisa},  with responses from 14,491 schools in 67 countries. The outcome variable, measured in ordinal scale, was the agreement with the statement ``In your school, to what extent is the learning of students hindered by the following phenomena? -- Teachers not meeting individual students' needs''. The question was answered by the principal/headteacher of the school, using the response categories ``Not at all'' (1), ``Very little'' (2), ``To some extent'' (3) or ``A lot'' (4). As independent variables, we included two measures for the size of the school, namely the average class size ($X_1$), recorded in categories 15 students or fewer, 16-20, 21-25, 26-30, 31-35, 36-40, 41-45, 46-50, more than 50 students, and the total enrollment in the school, recorded as a count ($X_2$). We excluded 69 schools where the reported average class size conflicted with the enrollment, thus including 14,422 in the analysis.

For our analysis, we used the empirical cumulative distribution function (ECDF) transformation to scale the covariates to the interval $[0,1],$ and then modeled their joint effect non-parametrically. The country effect was modeled with a logit link by including a zero mean normally distributed country-specific random effect into the linear predictor, with variance estimated from the data and updated in the MCMC algorithm from a conjugate inverse gamma posterior. The direction of monotonicity was fixed to be non-decreasing along both covariate axes. The non-parametric regression surfaces were allowed to vary on logit scale in the interval $[-5,5].$ The hyperparameters were chosen to be $a=0.1$ and $b=0.1$ for the Poisson point process intensities, with uniform priors for function levels ($d=0$). The MCMC algorithm was run for 10,000 rounds after a 5,000-round burn-in, saving every 20th iteration. Each round involved 3 death/birth and combined death-birth update proposals, and one proposal for all the other parameters. For a comparison, we also fitted a corresponding Bayesian PO mixed effects model, with additive log-linear effects for the two covariates on the logit scale (with total number of students entered into the model log transformed), which corresponds to the model specification \eqref{eq:ModelSemiparametric} without the $\mathbf x$ covariates, that is, $\lambda_k$s given just by the ordered intercept terms. We also fitted the corresponding GAM-type model \eqref{eq:ModelSemiparametric2} for the two covariate effects, where we specified 25 parameters corresponding to the equally spaced intervals on the ECDF transformed scale of the total enrollment, and a parameter for each level of the categorical class size variable. Random walk prior was assumed for these parameters, as described in Section \ref{sec:SemiparametricModel}

\begin{figure}[!h]
        \centering
        \subfloat[PISA data]{\label{loglika}\includegraphics[width=0.5\textwidth]{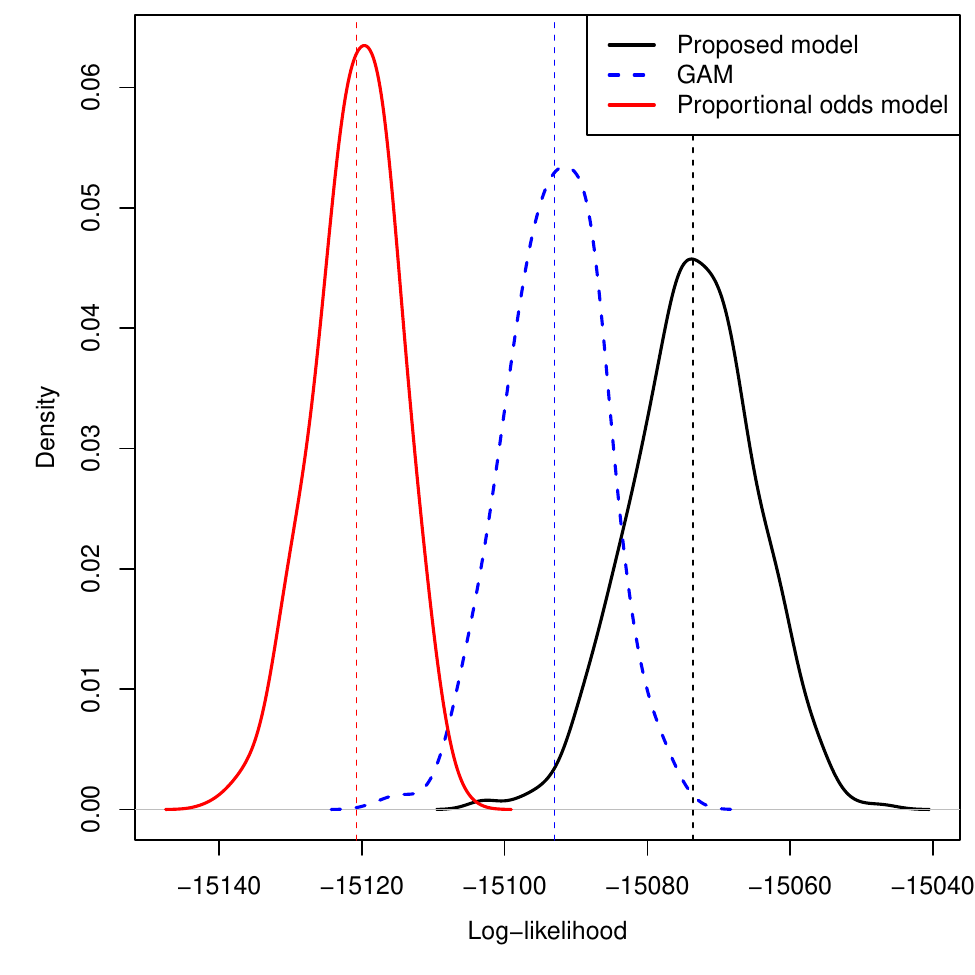}}
        \subfloat[Credit score data]{\label{loglikb}\includegraphics[width=0.5\textwidth]{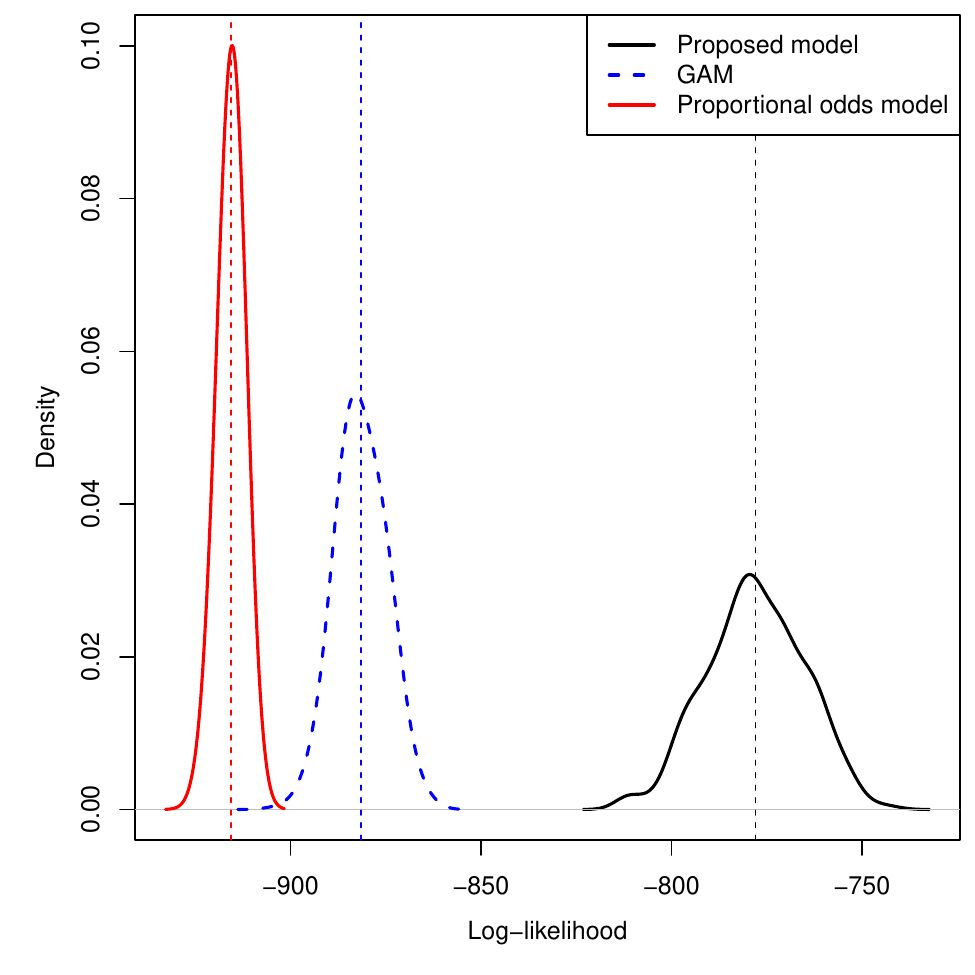}}
        \caption{Panel (a): Density plot of the log-likelihood posterior distribution for the proposed model (black line), GAM (dashed blue line), and the PO model (red line) fitted to PISA data. Panel (b): the same in the credit score data. The vertical dashed lines indicate posterior mean log-likelihood.}
\end{figure}

\begin{figure}[!h]
        \centering
        \includegraphics[width=\textwidth]{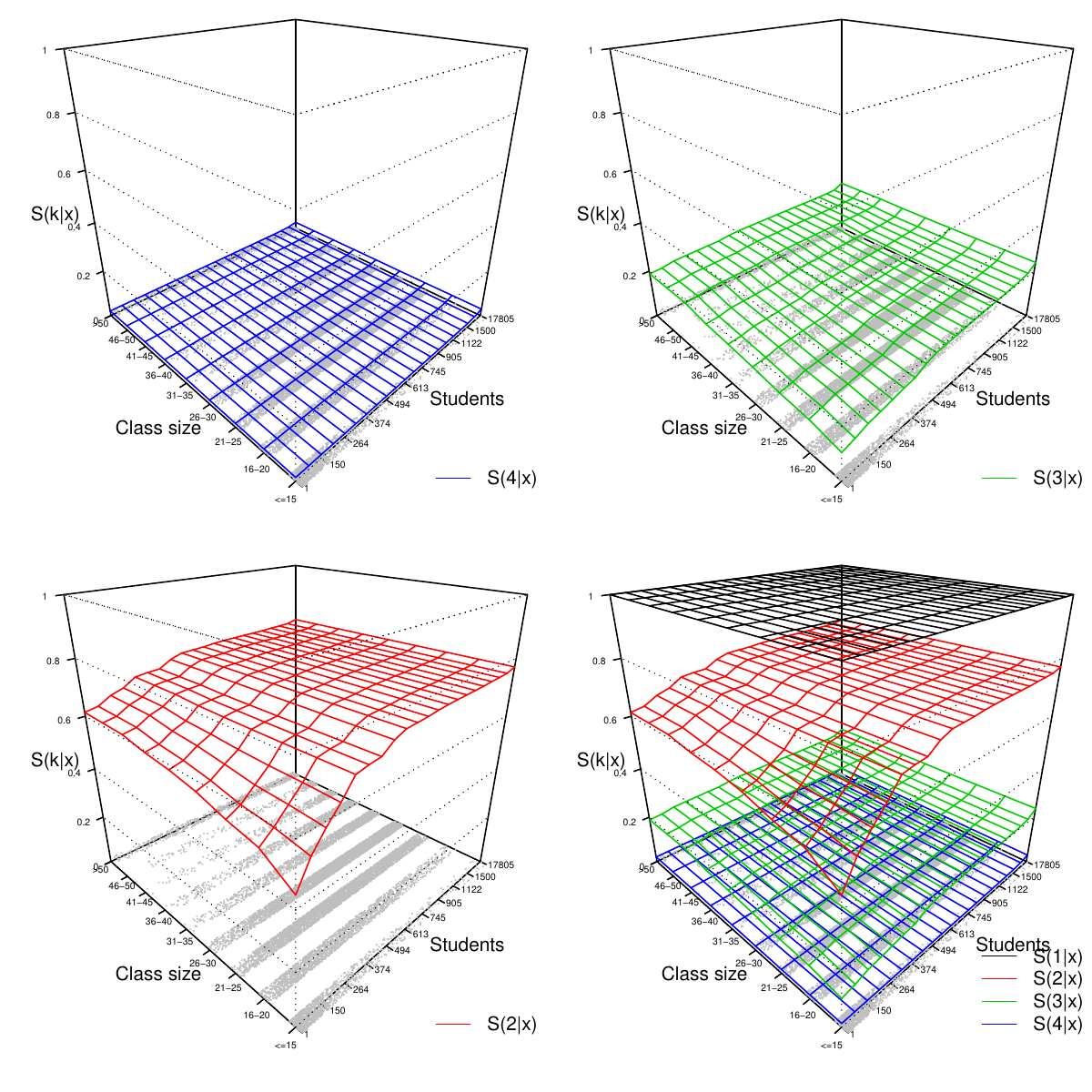}
        \caption{A perspective plot of the posterior mean regression surfaces for the three cumulative response category probabilities in the PISA data analysis. The jittered dots show the covariate coordinates of the data points.}\label{pisapersp}
\end{figure}

\begin{figure}[!h]
        \centering
        \includegraphics[width=\textwidth]{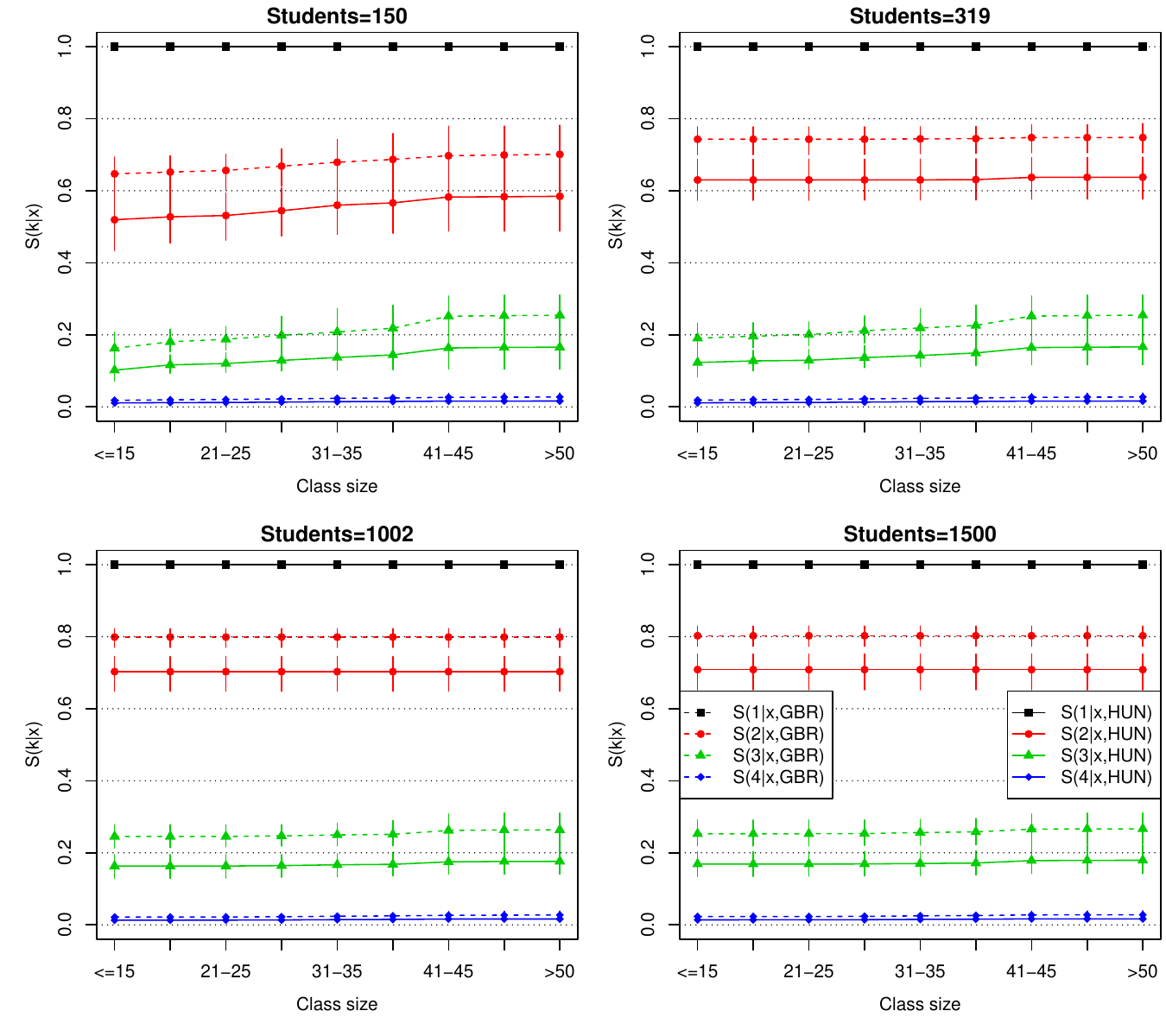}
        \caption{Conditional posterior mean regression functions for the three cumulative response category probabilities in the PISA data analysis, as functions of class size.}\label{pisacond}
\end{figure}

The 15,000 iterations took 6,176 seconds. The acceptance ratio for the birth, death, combined death-birth and position change moves were 52\%, 50\%, 23\% and 68\% for the point process $\Xi_1$ on the number of students axis, 46\%, 52\%, 30\% and 80\% for $\Xi_2$ on the class size axis, and 0.2\%, 73\%, 35\% and 47\% for the point process $\Xi_3$ on the plane for the two covariates. The regression surfaces were mainly constructed using the two one-dimensional point processes, with the posterior mean number of points in each process 13.2 ($\Xi_1$), 9.4 ($\Xi_2$) and 0.01 ($\Xi_3$). The acceptance ratio for the joint proposals for the level parameter vectors $\boldsymbol \delta_{ij}$ was 69\%, and the ratio for the single level change proposals for $\delta_{ijk}$, $k = 2, \ldots, 4$, higher than this, indicating that the prior proposals are highly informative 

The model fit for the three models is illustrated in Figure \ref{loglika}; this shows that the proposed monotonic model produced the best fit to the data, compared to the PO and GAM formulations. The GAM allows for flexible non-monotonic covariate effects while still assuming log-additivity and proportional odds, whereas the proposed model introduces the monotonicity assumption and relaxes both the additivity and PO assumptions. The resulting posterior mean regression surfaces for the three cumulative response category probabilities are shown in the perspective plot of Figure \ref{pisapersp}. The posterior mean regression surfaces were calculated at fixed covariate values $\mathbf x$ and cluster levels $c$ as $\frac{1}{L}\sum_{\ell=1}^L S(k \mid \mathbf x, c; \lambda_k^{(\ell)})$, where $\lambda_k^{(\ell)}, \ell=1,\ldots,L$ is the posterior sample for the level $k$ regression function. For Figure \ref{pisapersp}, the country effect was set to the expected value of zero, while the class size and school size were varied on a rectangular grid. The distribution of the data points on ground level shows the correlation between the two main covariates. Notable about the covariate effects is that the effect of the school size on the `not meeting the individual students' needs' response seems to level off, while the effect of the class size mainly seems to be present in relatively small schools, explaining the worse fit for the models assuming log-additive effects. For comparison, the posterior mean perspective plots for the GAM are shown in Supplementary Figure~S9.

There was a substantial country effect on the responses, with the posterior median random effect variance of 0.54 and 90\% credible interval $(0.39,0.72)$. Another way to present the model results is shown in Figure \ref{pisacond}, where the conditional posterior mean regression surfaces are presented as functions of class size at different values of school size. Such curves are shown for two countries, Great Britain (GBR) and Hungary (HUN), for which the 90\% credible intervals for the country specific random effects were in opposite directions, being then the first countries for which the effect did not overlap with zero (see the Supplementary Figure~S10 for all the country effects). The 90\% credible intervals for the regression surfaces are mostly non-overlapping for these two countries, and the country effect seems comparatively larger than either of the covariate effects.

\subsection{Credit score data set}\label{sec:credit}

In our second real data example, we wanted to demonstrate (i) the ability of the proposed method to incorporate multiple covariates, (ii) the workings of the inbuilt model selection functionality, (iii) the graphical presentation of the results in a higher dimensional setting, and (iv) data-adaptive choice of the direction of monotonicity, as proposed in Section \ref{sec:direction}. For this purpose, we reanalysed the credit score data set considered by \citet{chib2010additive} and \citet{deyoreo2018bayesian}. The response here was the 7-level credit rating of 921 US companies, re-coded into 5 categories by combining the two lowest and two highest ratings which were rare. The data set has five covariates, viz., book leverage (BOOKLEV, $X_1$), earnings before interest and taxes, divided by total assets (EBIT, $X_2$), log-sales (LOGSALES, $X_3$), retained earnings divided by total assets (RETA, $X_4$), and working capital divided by total assets (WKA, $X_5$). A priori, we could have assumed the direction of the monotonicity to be non-increasing for BOOKLEV and non-decreasing for EBIT, LOGSALES abd RETA. However, due to WKA being a ratio, the direction of this is more difficult to judge a priori. To demonstrate the feature of data-adaptive choice of the direction of monotonicity, we chose to leave this unspecified for all five covariates. All  covariate effects were modeled non-parametrically, allowing the model to reduce to lower dimensions by defining all 31 point processes corresponding to the non-empty subsets of the covariates.

We modeled the response with the logit link, allowing the surfaces to vary in the interval $[-10,10],$ but with no additional parametric restrictions. Otherwise, the priors were chosen as in the previous section. The covariates were re-scaled to the interval $[0,1]$ using the ECDF transformation. The model fit was compared to a reference Bayesian PO model with additive and linear effects on the logit scale, as well as to the GAM where each additive effect was modeled through 25 parameters corresponding to equally spaced intervals on the ECDF transformed scale of each of the five covariates. The MCMC algorithm was run for 10,000 rounds after a 5,000-round burn-in, saving every 20th iteration. Each iteration featured 31 birth/death and combined death/birth proposals, and a single proposal for all other parameters in the model.

Since in this context it might be of interest to consider each covariate's effect on the credit score when keeping the others constant, instead of the marginal regression functions, we present the results in terms of directly standardized regression functions for each covariate in turn, averaging over the empirical joint distributions of the other covariates as
\begin{equation*}
\frac{1}{L} \sum_{\ell=1}^L \frac{1}{N} \sum_{n=1}^N \textrm{expit}\left (\lambda_k^{(\ell)}(x_{1n}, \ldots, x_{j} = x, \ldots, x_{pn})\right)
\end{equation*}
for all $j, 1 \leq j \leq 5$. The resulting regression functions can then be presented as functions of $x$. While causal modeling is not the present focus, we note that under the assumption of absence of unmeasured confounding, these functions can be given an interpretation as `effects of $x$ on the credit score', holding the distribution of the other covariates constant. Similarly, if multivariable effects are of interest, we can calculate regression surfaces on a grid of values $(x,x')$ for covariates $j$ and $j'$ as 
\begin{equation*}
\frac{1}{L} \sum_{\ell=1}^L \frac{1}{N} \sum_{n=1}^N \textrm{expit}\left (\lambda_k^{(\ell)}(x_{1n}, \ldots, x_{j} = x, \ldots, x_{j'} = x', \ldots,  x_{pn})\right)
\end{equation*}
and present these averaged over the posterior distribution $\lambda_k$ in a 3-dimensional perspective plot.

 \begin{figure}
        \centering
        \includegraphics[width=\textwidth]{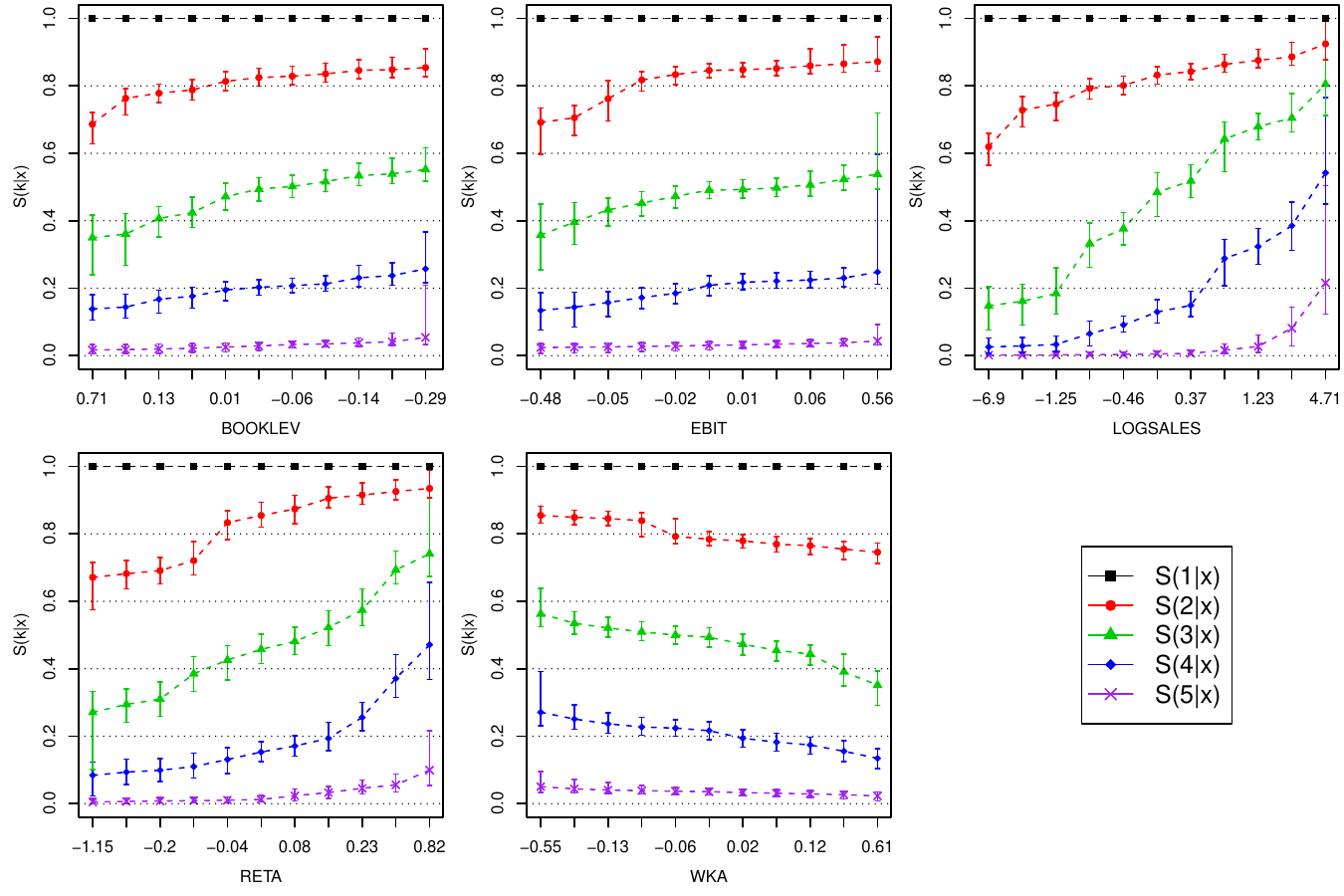}
        \caption{Standardized posterior mean regression functions for the four cumulative response category probabilities in the credit score data analysis, as functions of the different covariates.}\label{creditcumu}
\end{figure}

\begin{figure}[ht!]
\centering
\includegraphics[width=\textwidth]{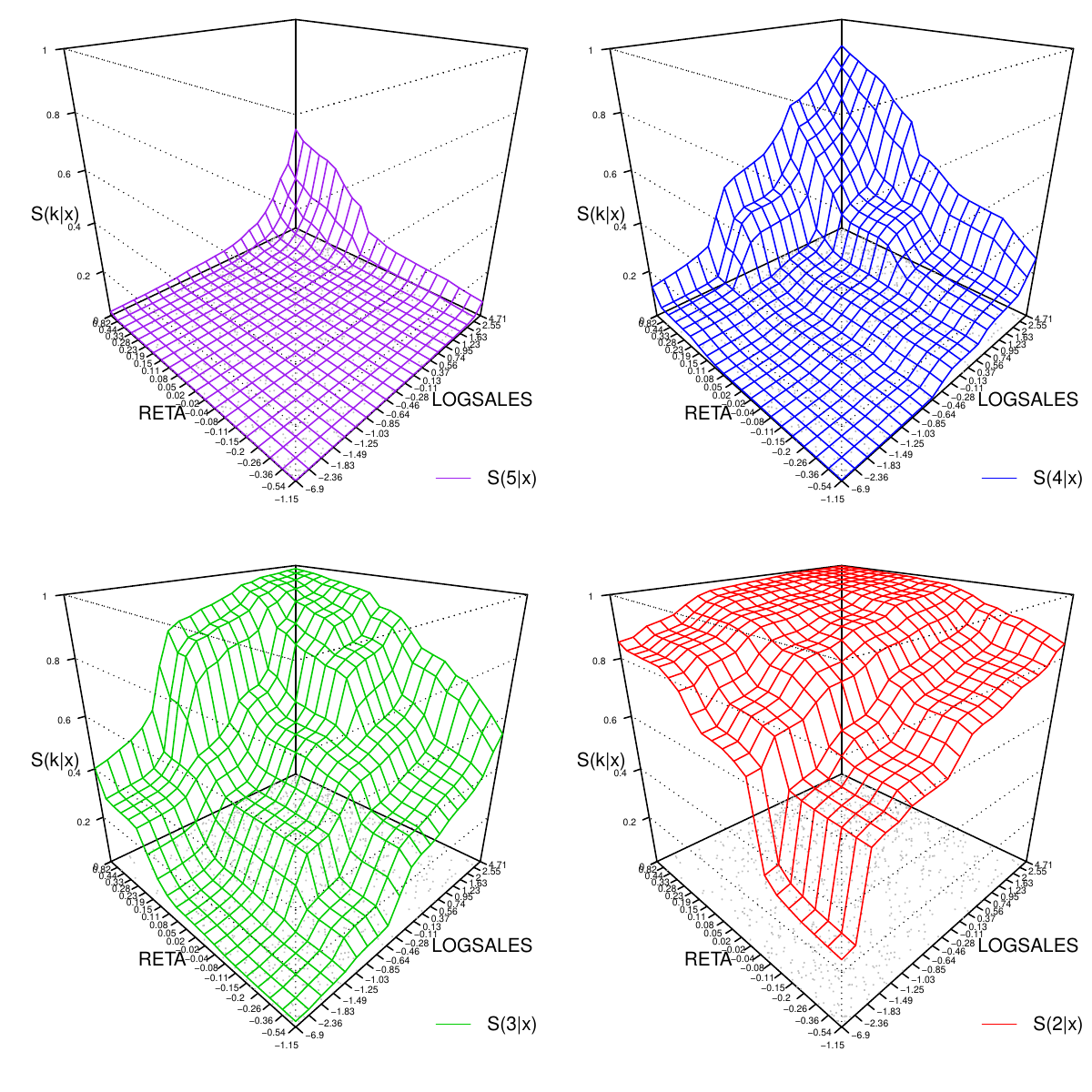}
\caption{Perspective plots of standardized posterior mean regression surfaces for the four cumulative response category probabilities as functions of covariates RETA and LOGSALES in the credit score data. The dots show the covariate coordinates of the data points.}
\label{credit3d}
\end{figure}

The 15,000 iterations took 7,898 seconds. The statistics for the 31 point processes in the model specification are presented in Supplementary Table~S1. They indicate that many different point processes were utilized to construct the regression surfaces, although the process with the largest number of points in the realizations was $\Xi_{30}$ specified in the space of the four covariates $X_2, X_3, X_4$ and $X_5$. Similar acceptance ratios for both birth and death proposals indicate that the process was consistently used in the model. In addition to the statistics in the table, the acceptance ratio for joint $\boldsymbol \delta_{ij}$ updates was 30\%, and the ratios for individual $\delta_{ijk}$, $k = 2, \ldots, 5$, updates ranged between 89\% and 94\%. 

The model fit comparison is presented in Figure \ref{loglikb}. We can see a significant improvement over the PO and GAM models, although there is considerable variability as the algorithm explores different model configurations. The standardized posterior regression functions are presented in Figure \ref{creditcumu}; they can be contrasted to the corresponding ones from the non-monotonic GAM, presented in Supplementary Figure~S11. The data-driven directions of monotonicity remained fixed after the burn-in run, corresponding to those initially hypothesized for $X_2, X_3, X_4$ and $X_5$, whereas it was reversed from such original guess for $X_5$ (WKA). LOGSALES and RETA emerged as the strongest predictors of credit rating. 

\citet{chib2010additive} reported some non-monotonicity in the covariate effects; however, our results with the GAM does not give real evidence for such a conclusion (Supplementary Figure~S11). In fact, the model fit comparison (Figure \ref{loglikb}) suggests that relaxing the additivity and proportional odds assumptions is more important in improving the fit. We did not contrast our results directly to \citet{deyoreo2018bayesian} as the standardized regression functions here are not directly comparable to the marginal regression functions presented therein. 

The probability of a covariate $j$ being included in the model can be calculated from the MCMC run by taking the proportion of the iterations where $n(\Delta_i) > 0$ for at least one of the point processes defined in a subset of covariates involving $j$. This probability was one for all the covariates. Another measure of covariate selection would be the total count of random points in the configurations for the point processes defined in a subset of covariates involving $j$. The posterior means for these were 18.6 (BOOKLEV), 20.9 (EBIT), 30.4 (LOGSALES), 28.6 (RETA) and 21.2 (WKA). We also run a model where the direction of monotonicity was fixed a priori for all the covariates and assumed to be non-decreasing for WKA. This effectively led to WKA being dropped from the model, with only 1.1 points on average used along this axis (results not shown).

An example of a 3-dimensional presentation of standardized regression surfaces, as functions of LOGSALES and RETA, is presented in Figure \ref{credit3d}, demonstrating a very strong association of the credit score with these two covariates, to the extent that, at certain combinations, some of the credit score levels are not present at all. 

\section{Discussion}\label{sec:discussion}

In this paper, we proposed a monotonic ordinal regression model and a Bayesian estimation procedure. The model differs from previous proposals in that it uses monotonicity as the only modeling assumption, without added proportionality, smoothness or distributional assumptions, while still enabling fully probabilistic inferences. Multivariable monotonicity of the covariate effects can thereby replace typically made stronger modeling assumptions such as additivity or linearity. A particular direction of monotonicity is often intuitively plausible on a priori basis; however, if uncertain, it can be left unspecified in the estimation algorithm. Our proposal then extends multivariable monotonic regression to ordered categorical outcomes, replacing the common proportional odds assumption with a stochastic monotonicity property of the ordered outcome categories with respect to the covariates, technically formulated here in terms of the corresponding cumulative probabilities. 
A general computational challenge in fitting non-proportional odds ordinal models is to ensure the ordering of the said cumulative probabilities; the present proposal resolves this in a natural way by combining the ordering constraints for monotonicity with the ordering of the regression functions for the cumulative probabilities of the outcome categories.

 As demonstrated in our simulation study, the proposed model construction based on marked point processes is flexible and can approximate different continuous and non-continuous monotonic regression surfaces with increasing precision when the sample size increases. However, because the piecewise constant realizations we used to construct the regression surfaces are relatively inefficient in approximating smooth functions, requiring a large number of support points for good approximation, the model may be best suited for large data sets. The advantage of the piecewise constant realizations is the ease of making local updating moves within the ordering constraints in the MCMC algorithm, but a consequence of lack of smoothness is the inability of the proposed model to extrapolate outside the support of the data. This may manifest as the ``spiking'' issue affecting traditional isotonic regression, leading to unstable estimates near the boundaries of the data. Due to the asymmetric nature of the proposed construction, we observed this phenomenon mainly close to the origin, where the regression surface may not be supported from below by data points, allowing it to drop abruptly. To counter this, we suggested a conditional prior specification that can borrow strength from the regression surface levels above. Specifying continuous or smooth regression function realizations based on the marked point process construction is a topic for further research. We note that alternatively to the marked point process prior, there are other prior processes that could possibly be adapted to the monotonic regression setting, such as Dirichlet process/stick-breaking priors; this, too, is a topic for further research.

Due to the curse of dimensionality, it would be unrealistic to model the effects of a very large number of covariates non-parametrically. Because of this, we proposed a built-in model selection feature that allows dropping redundant covariates from the model, as demonstrated in the application in Section \ref{sec:credit}. We also proposed a semi-parametric formulation that allows non-parametric modeling of the effects of the most important covariates, while simultaneously adjusting for a large number of other covariates. With this type of formulation, the PO assumption could be assessed by comparing models where a covariate is moved from the parametric to the non-parametric component; for example, by computing the corresponding Bayes factor. Similarly, the monotonicity assumption that we made throughout could  be tested in the Bayesian setting by comparing models with and without the constraint \citep{scott2015nonparametric}.

To compare our model specification to previously proposed formulations based on latent variables, briefly reviewed in Section \ref{section:intro}, we note that the basic formulation for the ordered outcome category specific survival probabilities $S(k \mid \mathbf x; \lambda_k) = \lambda_k(\mathbf x)$ simultaneously models the monotonic multivariable covariate effects and the distribution of the ordinal outcome variable in a flexible way. This remains the case if a monotonic link function is used, the only difference being that the uniform prior is then specified on the link function scale. Not requiring a continuous latent variable formulation may also have computational advantages; a separate data augmentation step is avoided since the likelihood can be evaluated directly given the current realizations of $\lambda_k(\mathbf x)$. Note also that, unlike the density regression type approaches, our approach does not involve modeling the distribution of the covariates.

As noted in Section \ref{section:intro}, latent variable formulations are well suited for extensions to multivariate ordinal outcomes, as the dependencies between the outcome variables can be captured by modeling the multivariate distribution of the latent variables. The present proposal may initially seem less suited for multivariate outcomes, but theoretically such extensions are possible. Consider, for example, the bivariate case of outcome variables $Y_1$ and $Y_2$ with, respectively, $K$ and $L$ ordered categories. A genuine stochastic ordering in two dimensions would be defined by taking (I) $P((Y_1, Y_2) \in \mathcal U \mid \mathbf x)$ to be non-decreasing in $\mathbf x$ for all upper sets $\mathcal U \subseteq \{1,\ldots,K\} \times \{1,\ldots, L\}$. A formally closer analogy to the univariate case would be to consider the bivariate survival functions $S(k,\ell \mid \mathbf x) = P(Y_1 \ge k, Y_2 \ge \ell \mid \mathbf x)$, $k=1,\ldots,K$, $\ell = 1,\ldots, L$, and then postulate the weaker monotonicity condition that (II) $S(k, \ell \mid \mathbf x)$ are non-decreasing in $\mathbf x$ for all $(k,l)$. A similar marked point process could be then specified to model the regression functions, with the points placed in the covariate space, and the marks in the space $[0,1]^{K \times L-1}$ of the survival probabilities. The survival functions themselves would follow the ordering $S(k,\ell \mid \mathbf x) \ge S(k',\ell' \mid \mathbf x)$ for all $1 \le k \le k' \le K$ and $1 \le \ell \le \ell' \le L$. The likelihood is constructed from the probabilities $P(Y_1=k,Y_2=\ell \mid \mathbf x) = S(k,\ell \mid \mathbf x) - S(k,\ell+1 \mid \mathbf x) - S(k+1,\ell \mid \mathbf x) + S(k+1,\ell+1 \mid \mathbf x)$, where $S$ would be additionally restricted by $0 \le P(Y_1=k,Y_2=\ell \mid \mathbf x) \le 1$ and $\sum_{k,\ell}P(Y_1=k,Y_2=\ell \mid \mathbf x) = 1$. Incorporating the stronger monotonicity condition (I) would require even more ordering constraints to be checked. Note that (I) holds if $P(Y_1 \ge k \mid \mathbf x)$ is non-decreasing in $\mathbf x$ for all $k$ and $P(Y_2 \ge \ell \mid \mathbf x, Y_1 = k)$ is non-decreasing in $\mathbf x$ and $k$ for all $\ell$. A similar sequence of one-dimensional conditions applies iteratively up to an arbitrary dimension.

We leave it to further work to study whether a computationally feasible implementation  of such a multivariate constructions can be found. Other potential extensions to be considered in future work are different shape constraints, for example, the `entire monotonicity' of \citet{fang2019multivariate}, which is a stronger restriction than the usual multivariable monotonicity, and partially ordered response categories in contrast to the linear ordering considered herein.  Finally, other spatial point processes than the homogeneous Poisson processes could potentially be used to construct the prior.

To conclude, we have proposed a non-parametric Bayesian regression model for ordered categorical outcome data and a corresponding estimation method. The estimation method is largely data driven, and it may require a large sample size to produce an accurate description of the regression surfaces. This is because, in  such data, individual data points carry only relatively little information about each specific level of the ordered outcome category. However, when large amounts of data are available, commonly made modeling assumptions such as linearity, additivity and proportional odds become restrictive and may lead to a poor fit. In that case, the multivariable monotonicity assumption provides a more flexible alternative.

\section*{Code and data availability}\label{sec:code}

The method has been implemented in the R package \texttt{monoreg} \citep{monoreg} distributed through CRAN (\url{https://CRAN.R-project.org/package=monoreg}). The datasets used for illustration are publicly available from OECD (\url{https://www.oecd.org/pisa/data/}) and from the Supplementary Materials to \citet{deyoreo2018bayesian}, available at \url{https://www.tandfonline.com/doi/suppl/10.1080/10618600.2017.1316280}.

\section*{Acknowledgement}\label{sec:acks}

The work of Olli Saarela was supported by a Discovery Grant from the Natural Sciences and Engineering Research Council of Canada. Christian Rohrbeck was beneficiary of an AXA Research Fund postdoctoral grant. The work of Elja Arjas was supported by the Big Insight research programme, University of Oslo. We are grateful to Arnoldo Frigessi for arranging for us a short visit at OCBE, during which some important steps were made towards completing this paper.

\bibliographystyle{apalike}
\bibliography{references}

\end{document}